# Second gradient poromechanics

Giulio Sciarra [a], Francesco dell'Isola [b,*], Olivier Coussy [c]

[a] *Dip. di Ingegneria Chimica, dei Materiali, delle Materie Prime e Metallurgia, Università di Roma "La Sapienza", via Eudossiana 18, 00184 Rome, Italy*
[b] *Dip. di Ingegneria Strutturale e Geotecnica, Università di Roma "La Sapienza", via Eudossiana 18, 00184 Rome, Italy*
[c] *Institut Navier ENPC, Av. Blaise Pascal, Cité Descartes, F77455 Marne-la-Vallée Cedex 2, France*

## Abstract

Second gradient theories have been developed in mechanics for treating different phenomena as capillarity in fluids, plasticity and friction in granular materials or shear band deformations. Here, there is an attempt of formulating a second gradient Biot like model for porous materials. In particular the interest is focused in describing the local dilatant behaviour of a porous material induced by pore opening elastic and capillary interaction phenomena among neighbouring pores and related micro-filtration phenomena by means of a continuum microstructured model. The main idea is to extend the classical macroscopic Biot model by including in the description second gradient effects. This is done by assuming that the surface contribution to the external work rate functional depends on the normal derivative of the velocity or equivalently assuming that the strain work rate functional depends on the porosity and strain gradients.

According to classical thermodynamics suitable restrictions for stresses and second gradient internal actions (hyperstresses) are recovered, so as to determine a suitable extended form of the constitutive relation and Darcy's law.

Finally a numerical application of the envisaged model to one-dimensional consolidation is developed; the obtained results generalize those by Terzaghi; in particular interesting phenomena occurring close to the consolidation external surface and the impermeable wall can be described, which were not accounted for previously.

*Keywords:* Second gradient materials; Porous media; Thermodynamics; Darcy law; Consolidation

## 1. Introduction

In standard continuum mechanics constitutive equations are usually formulated by assuming a local state postulate with regard to both time and space. When applied to fluids, it results in postulating that the (Helmholtz) free energy depends on the fluid mass density $\rho$ and on the temperature $T$ only, and not on the spatial gradients of the latter. Because only the first gradient of the fluid velocity is then involved when finally expressing the fluid mass balance, continuum mechanics based on the local state postulate is generally said to be a first gradient theory.

* Corresponding author.
 *E-mail address:* francesco.dellisola@uniroma1.it (F. dell'Isola).

First gradient mechanics reveals to be insufficient when trying to model the vapour–liquid transition via a continuum approach. Going back to the celebrated works of van der Waals, and restricting attention to a first gradient theory, the free energy density is known to be a non-convex function of the fluid mass density $\rho$. The minimizers of the energy, which actually govern the stable equilibrium states of the fluid during phase transition, are then restricted to the functions taking their values at the energy minima. In order to cure the bistable nature of the van der Waals model for phase transition, new information is mandatory. The missing information concerns the mechanical equilibrium of the liquid–vapour interface during phase transition. Since the mechanical equilibrium of the transition layer between the two co-existing phases is governed by non-local van der Waals forces, in a continuum approach to phase transition the free energy cannot depend on the local value of the mass density only. To be more precise, a first gradient model of phase transition is not capable of identifying the topological structure of the interface, which may be irregular or even dense. Accordingly a standard model to remove the bistable nature of a van der Waals-type approach to phase transition is the Cahn–Hilliard second gradient model (see Cahn and Hilliard, 1959). Second gradient theory introduces a dependence of the energy $\mathcal{E}$ on the density gradient according to

$$\mathcal{E}_\varepsilon = \int_\mathcal{D} \left[ W(\rho) + \varepsilon^2 |\nabla \rho|^2 \right] dv. \tag{1}$$

In this expression $W(\rho)$ is a double-well potential and $\varepsilon$ a small parameter that takes into account the energy that is connected with the formation of interfaces. As shown in Modica (1987a,b), as $\varepsilon \to 0^+$ the solution of the minimization problem associated to the functional (1) converges to the function $\rho_0$ which takes only the values $\alpha$ and $\beta$ corresponding to minima of the first gradient energy $W(\rho)$, while the set $E := \{x \in \mathcal{D} : \rho(x) = \alpha\}$ minimizes the surface area of its boundary among all subsets of $\mathcal{D}$ having the same volume. The pressure jump through the liquid/vapor interface is then governed by the usual Young–Laplace equation in terms of the product of surface tension and mean curvature (see de Gennes, 1985; Anderson et al., 1998). By contrast, when the problem at hand involves a characteristic length having the same order of magnitude as the thickness of the layer forming the liquid/vapor interface, the limit procedure $\varepsilon \to 0^+$ does not apply any more and the full procedure of the minimization of the functional (1) is required. This actually occurs for phenomena involving thin liquid films (Seppecher, 1993), droplets exhibiting small curvature radius (dell'Isola et al., 1996; Gavrilyuk and Saurel, 2002), or a topological transition associated for instance to pinching and fissioning of liquid jets (Lowengrub and Truskinovsky, 1998). In describing these phenomena more refined models than those based on the simple Young–Laplace equation are needed.

This kind of problem has been discussed in the literature both for the case of a single-component fluid and binary fluids according to a macroscopic approach (see e.g. Anderson et al., 1998; Seppecher, 1996; Lowengrub and Truskinovsky, 1998; Lee et al., 2002) considering the compressible, incompressible and quasi-incompressible behaviour (see Lowengrub and Truskinovsky, 1998 for a detailed discussion) of the bulk material. It is worth noticing that quasi-incompressibility means that the overall mass density is not constant, even if mixture constituents are incompressible, because of a non-constant constituent mass concentration; in this case the set of balance laws is enriched by a mass concentration equation (see Lowengrub and Truskinovsky, 1998) or a micro-force balance law (see Gurtin et al., 1996). For multiphase systems, as mixtures of non-interacting fluids, a second gradient model involving co-capillarity can also be introduced (see e.g. Seppecher, 1987) which is capable of describing the effect of compressibility of the constituents on the behaviour of the mixture as a whole.

Early formulations of second gradient mechanics of solids (Germain, 1973) consist of considering both the first and the second gradient of the displacement, or equivalently the deformation and its spatial gradient, as relevant candidates in the expression of the deformation working. Extended second gradient mechanics of solids consists of considering both the first and the second gradient of relevant state variables in the expression of the free energy.

Following the pioneering work of Biot in poroelasticity (Biot, 1941), poromechanics (Coussy, 2004) is that branch of mechanics which deals with the behaviour of deformable porous solids whose internal solid walls are subjected to the pressure of an interstitial fluid. Once a microscopic description of the porous material is introduced, it is generally accepted that the macroscopic model can be viewed as a suitable average of the microscopic one. This requires to identify every reference volume element (RVE) of macroscopic continuum mechanics with its centroid (Dormieux and Ulm, 2005). The up-scaling procedure is worked out by assuming uniform strain (stress) boundary conditions for every reference volume in terms of the macroscopic strain (stress), or periodic boundary conditions. Obviously the microscopic strain (stress) distribution may not be spatially uniform in the reference element volume because of the included heterogeneities. Therefore there is no chance to recover continuity of the microscopic stress vector on the interface between two contiguous reference element volumes using the standard first gradient theory. As a result a macroscopic first gradient theory is not capable to address the effects associated to neighbouring pores of contiguous reference element volumes. Following Drugan and Willis (1996), one can induce the characteristic size of the reference volume element by requiring the non-local corrections to the constitutive law to be within a given error percentage of the standard effective stiffness. In addition to be highly heterogeneous, liquid-saturated porous solids have fluid–solid interfaces whose thickness may become comparable with the pore size. These are the conceptual reasons for exploring higher gradient poromechanics.

Second gradient mechanics for porous solids has actually been already addressed in the framework of the theory of mixtures in dell'Isola et al. (2000), Sciarra et al. (2001) and dell'Isola et al. (2003) and recently in that of poromechanics (see Collin et al., 2006) by considering a second gradient behaviour for the solid only. The approach developed here extends these results by providing a macroscopic Biot-like theory of poroelastic materials for which second gradient effects are associated to both solid and fluid constituents. With respect to the results presented in Collin et al. (2006), here a complete deduction of second gradient poromechanics has been developed, together with an explicit formulation of a constitutive macroscopic model, consistent with thermodynamics.

In Section 2, some preliminaries are discussed in order to make clear the following developments. In Section 3, we extend first gradient poromechanics (Coussy, 2004) to second gradient poromechanics. The approach allows us to express the internal work rate with the help of additional stress fields compatible with the external work rate accounting for second gradient effects. In the following sections the second gradient poroelastic constitutive equations and modified Darcy's law are derived from thermodynamics. In the last section, a numerical example is developed for treating the classical one-dimensional consolidation problem stated by von Terzaghi (1946) in the framework of the second gradient poromechanical model. The new model is capable to cure the singular behaviour of Terzaghi's solution close to the surface on which the consolidating loading is exerted. The analysis reveals a Mandel-like dilatant effect (Mandel, 1953) in the early consolidation process close to the impermeable wall.

## 2. Preliminaries and notations

In this section, we introduce some formalism to be adopted in the following. We in particular point out the distinction between Eulerian and Lagrangean fields and the way in which the ones transform into the others. According with classical poromechanics, we introduce the Lagrangean field equations starting from the Eulerian ones, deduced from the Principle of Virtual Powers. It is worth to notice that in the Lagrangean description of motion all the fields in the governing equations are regarded as dependent on the reference placement in the initial configuration of the solid; on the other hand in the Eulerian description the governing equations are prescribed in terms of spatial fields (i.e. fields defined over the current configuration).

Let $\chi(\cdot, t)$ indicate the placement of a solid material particle $X_s$. This is a diffeomorphism over the Eucledean place manifold $\mathcal{E}$, such that $x = \chi(X_s, t)$ represents the current placement of $X_s$. $\mathbf{F}(\cdot, t)$ indicates its gradient. In this paper, we explicitly distinguish between the Lagrangean gradient ($\nabla_0$), with respect to the reference place in the configuration of the solid skeleton, and the Eulerian gradient ($\nabla$), with respect to the current place $x$. Analogously, the solid Lagrangean and the Eulerian divergence operators will be noted by $\text{div}_0$ and $\text{div}$, respectively. All the classical transport formulas can be derived; in particular those ones for a material volume and for an oriented surface element turn to be

$$\mathrm{d}v = J\,\mathrm{d}v_0 \qquad \mathbf{n}\,\mathrm{d}s = J\,\mathbf{F}^{-T}\mathbf{n}_0\,\mathrm{d}s_0 \tag{2}$$

where $\mathrm{d}v$ and $\mathrm{d}s$ represent the current elementary volume and elementary oriented surface corresponding to the reference ones $\mathrm{d}v_0$ and $\mathrm{d}s_0$, moreover $J = \det \mathbf{F}$. In the following $\mathcal{D}$ will indicate the current domain occupied by the porous material and $\mathcal{D}_0$ its solid reference shape. According with the previous remarks the following rules for transforming Eulerian into Lagrangean gradient and divergence for a spatial vector (second order tensor) field **v** (**V**) hold true (for more details see e.g. Truesdell, 1991; Gurtin, 1981):

$$\begin{cases} \nabla \mathbf{v} = \nabla_0 (\mathbf{v} \circ \chi) \mathbf{F}^{-1}, \\ J \mathrm{div} \mathbf{v} = \mathrm{div}_0 [J \mathbf{F}^{-1} (\mathbf{v} \circ \chi)] \end{cases} \quad \begin{cases} \nabla \mathbf{V} = \nabla_0 (\mathbf{V} \circ \chi) \mathbf{F}^{-1}, \\ J \mathrm{div} \mathbf{V} = \mathrm{div}_0 [J (\mathbf{V} \circ \chi) \mathbf{F}^{-T}], \end{cases} \tag{3}$$

where $\mathbf{v} \circ \chi$ and $\mathbf{V} \circ \chi$ are the Lagrangean fields associated with the spatial **v** and **V** by means of the solid placement map. In this paper overall, as well as intrinsic fluid, stresses and hyper-stresses (these last to account for second gradient contributions) are introduced as Eulerian fields ($\mathbf{T}, \mathbf{T}_f, \mathbb{C}, \mathbb{C}_f$); however their Lagrangean pull back, in the reference configuration of the skeleton, will be the only stress fields constitutively prescribed in terms of a suitable macroscopic potential (see Section 5). We use identities (3) for introducing, at the end of Section 3, the relation between the Cauchy stress tensors (**T** and $\mathbf{T}_f$) and the Piola–Kirchhoff stress tensors (S and $\mathbb{S}_f$) for the overall porous material and the pure fluid, as well as the relation between the Eulerian (**c** and $\mathbf{c}_f$) and Lagrangean ($\gamma$ and $\gamma_f$) hyper-stress vectors, which univocally prescribe hyper-stresses with some rescriting assumptions on admissible second gradient tractions (see Eq. 18).

## 3. Second gradient poromechanics

We start by recalling second gradient mechanics. For a monophasic continuum referred to by index *a*, following Toupin (1962, 1964), Mindlin (1964), Casal (1972) and Germain (1973) the external working $\mathcal{W}^{\mathrm{ext}}$, which accounts for second gradient effects, is a continuous linear functional of the field velocity $\mathbf{v}_a$ (see e.g. dell'Isola and Seppecher, 1997; Degiovanni et al., 2006) such as

$$\mathcal{W}^{\mathrm{ext}}(\mathbf{v}_a) = \int_{\mathcal{D}} \mathbf{b}_a \cdot \mathbf{v}_a \, dv + \int_{\partial \mathcal{D}} \left( \mathbf{t}_a \cdot \mathbf{v}_a + \boldsymbol{\tau}_a \cdot \frac{\partial \mathbf{v}_a}{\partial \mathbf{n}} \right) ds + \sum_{i=1}^{n_e} \int_{\mathcal{L}_i} \mathbf{t}_a^i \cdot \mathbf{v}_a \, dl, \tag{4}$$

where $\mathcal{D}$ is the current domain (or any of its regular subdomains), $\partial \mathcal{D}$ its boundary and $n_e$ the number of edges $\mathcal{L}_i$ of the boundary, if any. Moreover $\mathbf{b}_a$ is the density of body forces, $\mathbf{t}_a$ the surface traction, $\boldsymbol{\tau}_a$ the so called double-force (or "1-normal" contact forces) and $\mathbf{t}_a^i$ the edge force per unit line acting along the *i*th edge of the boundary. Note that $\partial/\partial \mathbf{n}$ is the normal derivative, that is the directional derivative along the outward unit normal. According to the primary results on second gradient theories stated by Germain (1973) the external action $\boldsymbol{\tau}_a$ can be regarded as the sum of two different contributions: (i) an external areal double force working on the stretching velocity, ($\nabla \mathbf{v}_a \cdot \mathbf{n} \otimes \mathbf{n}$) along the outward unit normal **n** of the boundary; (ii) a tangential couple working on the vorticity. If in particular only the vorticity effect was considered then the envisaged model could be regarded as a kind of Cosserat continuum where local rotations are prescribed in terms of the velocity gradients (see e.g. Suiker et al., 2001). Consistently with (4), following Germain (1973) the strain working related to second gradient mechanics is now expressed in the form[1]

$$\mathcal{W}^{\mathrm{int}}(\mathbf{v}_a) = \int_{\mathcal{D}} [\mathfrak{T}_a \cdot \nabla \mathbf{v}_a + \mathbb{C}_a \cdot \nabla \nabla \mathbf{v}_a] \, dv, \tag{5}$$

where $\mathfrak{T}_a$ and $\mathbb{C}_a$ stand, respectively, for the stress tensor and for the hyper-stress tensor. For the sake of simplicity, ignoring inertia forces, the Principle of Virtual Power states that $\mathcal{W}^{\mathrm{ext}}(\mathbf{v}_a) = \mathcal{W}^{\mathrm{int}}(\mathbf{v}_a)$, whatever the velocity field $\mathbf{v}_a$. Exploiting this equality, standard procedures lead to the momentum equation

$$\mathrm{div}(\mathfrak{T}_a - \mathrm{div}\mathbb{C}_a) + \mathbf{b}_a = 0 \tag{6}$$

and the conditions on the boundary of every regular subdomain of $\mathcal{D}$

$$\mathbf{t}_a = [\mathfrak{T}_a - \mathrm{div}(\mathbb{C}_a)]\mathbf{n} - \mathrm{div}_S[\mathbb{C}_a \mathbf{n}]; \quad \boldsymbol{\tau}_a = (\mathbb{C}_a \mathbf{n})\mathbf{n}, \tag{7}$$

and on everyone of its edges $i = 1, \ldots, n_e$

$$\mathbf{t}_a^i = [\![ (\mathbb{C}_a \mathbf{n}) \nu ]\!]_i, \tag{8}$$

---

[1] From now on the central dot will indicate the inner product between two *n*th order tensors.

where $\text{div}_S$ and $[\![\varphi]\!]_i$ stand, respectively, for the surface divergence over the regular parts of the boundary, and for the jump of $\varphi$ across the $i$th edge, respectively. Eq. (7) can be reformulated in order to underline how (external) tractions exerted on the regular parts of the boundary depend on the contact surface not only via a linear map acting on the normal unit vector **n** components, but also via a linear combination of the components of the curvature tensor $\mathbf{N} = \nabla_S \mathbf{n}$:[2]

$$t_a = \{\mathfrak{T}_a - \text{div}\mathfrak{C}_a - \text{div}_S(\mathfrak{C}_a^\top)^\top\}\mathbf{n} - \mathfrak{C}_a \mathbf{N}. \tag{9}$$

Let us now consider a solid–fluid mixture and refer to the solid and fluid constituents by the indices s and f, respectively. At any time a solid–fluid mixture can be viewed as the superposition of two interacting continua, $a = $ s and $a = $ f. Accordingly, the Principle of Virtual Power, and consequently (6)–(8), apply separately to the solid continuum and to the fluid continuum. In addition $\mathfrak{T}_a$ and $\mathfrak{C}_a$ are tensorial volumetric densities with regard to the macroscopic overall volume $dv$. We may re-express $\mathfrak{T}_a$ and $\mathfrak{C}_a$ by introducing tensorial densities $\mathbf{T}_a$ and $\mathbb{C}_a$ intrinsic to the constituents by referring them to the volume which the two constituents actually occupy. If $n$ is the Eulerian porosity, then $n\,dv$ is the current porous volume and, in the case of saturation, we can write

$$\mathfrak{T}_s = (1-n)\mathbf{T}_s; \quad \mathfrak{C}_s = (1-n)\mathbb{C}_s; \quad \mathfrak{T}_f = n\mathbf{T}_f; \quad \mathfrak{C}_f = n\mathbb{C}_f. \tag{10}$$

Accordingly, from (6) we finally obtain

$$\text{div}((1-n)\mathbf{T}_s - \text{div}[(1-n)\mathbb{C}_s]) + \mathbf{b}_s = 0; \quad \text{div}((n\mathbf{T}_f) - \text{div}(n\mathbb{C}_f)) + \mathbf{b}_f = 0. \tag{11}$$

Furthermore the body force $\mathbf{b}_a$ can be split in two terms according to

$$\mathbf{b}_s = \mathbf{b}_s + \mathbf{b}_{f \to s} = \mathbf{b}_s - \mathbf{b}_{s \to f}; \quad \mathbf{b}_f = \mathbf{b}_f + \mathbf{b}_{s \to f} = \mathbf{b}_f - \mathbf{b}_{f \to s} \tag{12}$$

where $\mathbf{b}_{f \to s}$ (resp. $\mathbf{b}_{s \to f}$) is the interaction force exerted by the fluid (resp. solid) continuum on the solid (resp. fluid) continuum. As indicated in (12), by virtue of the action and reaction law, we have $\mathbf{b}_{f \to s} = -\mathbf{b}_{s \to f}$, as it can be directly recovered from the Principle of Virtual Power (Dormieux et al., 1991). Adding the two equations obtained by successively letting $a = $ s and $a = $ f in (6) we then derive

$$\text{div}(\mathbf{T} - \text{div}\mathbb{C}) + \mathbf{b} = 0, \tag{13}$$

where **T** and $\mathbb{C}$ are, respectively, the overall stress tensor and the overall hyper-stress tensor:

$$\mathbf{T} = (1-n)\mathbf{T}_s + n\mathbf{T}_f, \tag{14}$$
$$\mathbb{C} = (1-n)\mathbb{C}_s + n\mathbb{C}_f. \tag{15}$$

The external working $\mathcal{W}_{\text{mix}}^{\text{ext}}(\mathbf{v}_s, \mathbf{v}_f)$ related to the mixture is the sum of the working related to each constituent. We write

$$\mathcal{W}_{\text{mix}}^{\text{ext}}(\mathbf{v}_s, \mathbf{v}_f) = \mathcal{W}^{\text{ext}}(\mathbf{v}_s) + \mathcal{W}^{\text{ext}}(\mathbf{v}_f). \tag{16}$$

The deformation working $\mathcal{W}_{\text{mix}}^{\text{int}}$ related to the mixture is obtained through exploring the equality $\mathcal{W}_{\text{mix}}^{\text{ext}} = \mathcal{W}_{\text{mix}}^{\text{int}}$. With the help of $(11)_2$ and of $(12)_2$, by eliminating the unknown interaction force $\mathbf{b}_{f \to s}$ from (16), we finally obtain after some straightforward mathematical manipulations

$$\mathcal{W}_{\text{mix}}^{\text{int}}(\mathbf{v}, \mathbf{w}) = \int_{\mathcal{D}} [\mathbf{T} \cdot \nabla \mathbf{v} + \text{div}(n\mathbf{T}_f^\top \mathbf{w}) + \mathbb{C} \cdot \nabla\nabla \mathbf{v} + n\mathbb{C}_f \cdot \nabla\nabla \mathbf{w} - \text{div}(\text{div}\, n\mathbb{C}_f) \cdot \mathbf{w}]\,dv, \tag{17}$$

where, and from now on, we use the notation $\mathbf{v} = \mathbf{v}_s$ and $\mathbf{w} = \mathbf{v}_f - \mathbf{v}_s$ in view of the forthcoming Lagrangean description with regard to the solid skeleton. For the sake of simplicity, the contribution of external bulk actions $\mathbf{b}_a$ has not been considered in (17). The two first terms are the ones related to a first gradient approach (Coussy, 2004), while the remaining terms capture the second gradient effects.

---

[2] In the following formulas the transpose of a third order tensor $\mathbb{C}$ is evaluated so as to fulfill the identity:

$$(\mathbb{C} \cdot \mathbf{v}) : \mathbf{V} = (\mathbb{C}^\top : \mathbf{V}) \cdot \mathbf{v}, \quad \forall \mathbf{v} \in \mathbb{R}^n \text{ and } \mathbf{V} \in \text{LIN}(\mathbb{R}^n).$$

From now on we restrict our attention only to those second gradient effects that are related to double forces, or from the kinematical point of view, to stretching velocities; in other words we assume skew-symmetric couples working on vorticity on the boundary to be negligible for both the solid and the fluid constituents. From a physical point of view such an assumption implies both solid granular materials and complex rheological fluids not to be modeled by the addressed second gradient formulation; on the other hand solid local dilatancies, capillarity of the fluid as well as co-capillarity between the solid and the fluid are going to be taken into account by the envisaged model (a more detailed discussion of these topics can be found in Section 5). The formulated hypothesis implies the hyper-stress tensor of both constituents to reduce to the form

$$\mathbb{C}_a = \mathbf{I} \otimes \mathbf{c}_a, \quad a = s, f, \tag{18}$$

where $\mathbf{I}$ is the second order identity tensor and $\mathbf{c}_\alpha$ a suitable vector field. According to Eq. (4), the second gradient external working is

$$\boldsymbol{\tau}_a \cdot \frac{\partial \mathbf{v}_a}{\partial \mathbf{n}} = n_a[(\mathbf{I} \otimes \mathbf{c}_a)\mathbf{n}]\mathbf{n} \cdot \frac{\partial \mathbf{v}_a}{\partial \mathbf{n}} = n_a(\mathbf{c}_a \cdot \mathbf{n})\mathbf{n} \cdot \frac{\partial \mathbf{v}_a}{\partial \mathbf{n}} = n_a(\mathbf{c}_a \cdot \mathbf{n})(\nabla \mathbf{v}_a \cdot \mathbf{n} \otimes \mathbf{n}), \quad n_a = \begin{cases} 1-n, & a=s, \\ n, & a=f. \end{cases} \tag{19}$$

As previously noticed $(\nabla \mathbf{v}_a \cdot \mathbf{n} \otimes \mathbf{n})$ represents the stretching velocity, thus $n_a(\mathbf{c}_a \cdot \mathbf{n})$ indicates the so called doubly normal double force acting on the $a$th constituent (see Germain, 1973).

Consider now a Lagrangean description of the porous material, where the initial reference configuration is that of the solid skeleton; the internal (generalized) stresses of the solid are therefore pulled back to this initial configuration via the following relations:

$$\mathbf{T} \cdot \nabla \mathbf{v} dv = \mathbb{S} \cdot \frac{d\boldsymbol{\Delta}}{dt} dv_0, \quad \mathbb{S} := J\mathbf{F}^{-1}\mathbf{T}\mathbf{F}^{-\top}, \tag{20}$$

$$\mathbb{C} \cdot \nabla \nabla \mathbf{v} dv = (\mathbf{I} \otimes \mathbf{c} \cdot \nabla \nabla \mathbf{v}) dv = \mathbf{c} \cdot \nabla(\mathrm{div}\mathbf{v}) dv$$
$$= \gamma \cdot \left[ \left(\nabla_0 \frac{d\boldsymbol{\Delta}}{dt}\right)^\top \mathbf{C}^{-1} - (\nabla_0 \mathbf{C})^\top \mathbf{C}^{-1} \frac{d\boldsymbol{\Delta}}{dt} \mathbf{C}^{-1} \right] dv_0, \quad \gamma := J\mathbf{F}^{-1}\mathbf{c} \tag{21}$$

we remind that $\mathbf{F}$ is the spatial gradient of the placement map of the solid, $\mathbf{C} := \mathbf{F}^\top \mathbf{F}$, $\boldsymbol{\Delta} := \frac{1}{2}(\mathbf{C} - \mathbf{I})$ and $J = \det \mathbf{F}$ the Cauchy–Green strain tensor, the Green–Lagrange strain tensor and the Jacobian of $\mathbf{F}$, respectively. Moreover $dv_0$ represents the reference volume element such that $dv = Jdv_0$ (see Section 2). The Lagrangean overall stresses are represented by $\mathbb{S}$ and $\gamma$, which indicate the second Piola–Kirchhoff stress tensor and the hyper-stress vector. In the previous formulas $\nabla$ indicates the Eulerian gradient and $\nabla_0$ the corresponding Lagrangean gradient. When introducing also the solid Lagrangean pull back of the fluid stress and hyper-stress

$$\mathbb{S}_f := J\mathbf{F}^{-1}\mathbf{T}_f \mathbf{F}^{-\top}, \quad \gamma_f := J\mathbf{F}^{-1}\mathbf{c}_f, \tag{22}$$

the internal deformation working takes, finally, the following form

$$\mathcal{W}_\mathcal{L}^{\mathrm{int}} = \int_{\mathcal{D}_0} \left\{ [\mathbb{S} - \mathbf{C}^{-1}((\nabla_0 \mathbf{C})\gamma)\mathbf{C}^{-1}] \cdot \frac{d\boldsymbol{\Delta}}{dt} + \mathbf{C}^{-1} \otimes \gamma \cdot \nabla_0 \frac{d\boldsymbol{\Delta}}{dt} + \mathrm{div}_0\left(\frac{1}{J\rho_f} \mathbb{S}_f^\top \mathbf{M}\right) \right.$$
$$\left. + \nabla_0 \left[J^{-1}\mathrm{div}_0\left(\frac{1}{n\rho_f}\mathbf{M}\right)\right] \cdot n\gamma_f - \frac{1}{n\rho_f}\mathbf{M} \cdot \nabla_0[J^{-1}\mathrm{div}_0(n\gamma_f)] \right\} dv_0, \tag{23}$$

where $\mathbf{M} := m_f \mathbf{F}^{-1}\mathbf{w}$ is the Lagrangean filtration vector and $m_f := Jn\rho_f$ the Lagrangean fluid mass content.

## 4. Thermodynamics of second gradient porous media

In this section, the first and the second principle of thermodynamics for the case of fluid saturated second gradient porous media will be discussed. According to the standard procedure based on the Clausius–Duhem inequality we extend the well known results due to Biot (see e.g. Biot, 1941) and determine, following Coussy et al. (1998) and Coussy (2004), a suitable macro-scale potential so as to identify for every state parameter the corresponding conjugate thermodynamic state variable. Such a potential will be the overall Helmholtz free energy. Moreover, prescribing a priori the specific Helmholtz free energy of the fluid, a suitable macro-scale potential just for the solid skeleton will be obtained by removing the free energy of the fluid from that of the whole body. In this case some special restrictions will be implied for the envisaged second gradient contribution to the strain energy.

Assume the overall internal energy to be prescribed in the form

$$e = \rho_s(1-n)e_s + \rho_f n e_f, \tag{24}$$

where $e_a$ ($a = s,f$) is the specific internal energy of the $a$th constituent and, according to the hypothesis of saturation, the volume fraction occupied by the solid is prescribed in terms of the porosity $n$. It is possible to prove that Eq. (24) represents the overall internal energy of a first gradient solid–fluid mixture, where the energy of the solid depends both on solid strain and porosity, whilst the energy of the fluid only depends on the density of the pure fluid (see e.g. Sciarra, 2001). In a second gradient mixture there is no chance for the specific internal energy of the $a$th constituent to depend only on the corresponding specific kinematical parameter (see e.g. Seppecher, 1987). This means that a dependence of the $a$th specific internal energy on the gradient of the corresponding apparent field[3] should be taken into account. In other words modeling topological transitions by considering the interfaces as narrow transition layers, where the constituents mix, does not allow to neglect porosity gradients in the overall internal energy (see Lowengrub and Truskinovsky, 1998).

Let us focus attention on the behaviour of the fluid and assume that its constitutive characterization is known in case of a second gradient description. We aim at capturing the expression for the free energy of the skeleton by removing the contribution only due to the fluid constituent from the overall Helmholtz functional. This procedure will allow for the determination of the second gradient extended formula for the macro-scale potential of the skeleton with respect to the classical one of the Biot model. We assume the specific energy of the fluid to be a function of $s_f$, $1/\rho_f$ and $\nabla \rho_f$, $s_f$ being the specific entropy of the fluid; however, according to the aforementioned remarks, we consider that only gradients of the fluid apparent density are liable to appear in the specific Helmholtz free energy of the fluid. This means that porosity and porosity gradient should be regarded as architectural parameters and can not be considered as state variables: they prescribe where the fluid must be confined, at every time step, but they do not affect the variation of what we claim to label the internal energy of the fluid. When evaluating the variation of the internal energy of the fluid, which fills the pore space, we therefore are going to regard the Eulerian porosity $n$ and its gradient $\nabla n$ as constants. Apparently, variation of porosity gradient should affect the remaining skeleton potential.

Finally when removing the pure-fluid-energy contribution from the overall free energy $\Psi$ or, in the same way, when incorporating the variation of the free energy with respect to porosity just in the variation of the energy associated to the solid skeleton, we do not regard $n$ and $\nabla n$ as state variables affecting the internal energy $e_f$ (Helmholtz free energy $\psi_f$) of the fluid.

Using a niggling notation we shall label in the following this variation $d_n$ and in particular for the internal energy of the fluid $e_f$ we get

$$d_n e_f := de_f|_{(n,\nabla n)=\text{const}} = \frac{\partial e_f}{\partial(1/\rho_f)} d\left(\frac{1}{\rho_f}\right) + n\frac{\partial e_f}{\partial \nabla(n\rho_f)} \cdot d\nabla \rho_f + \frac{\partial e_f}{\partial s_f} ds_f. \tag{25}$$

As we shall see in Section 4.2 this assumption guarantees the Clausius–Duhem inequality to be written in terms of time derivatives of the macro fields $\dot{\mathbf{\Lambda}}$, $\nabla_0 \dot{\mathbf{\Lambda}}$, $\dot{\phi}$, $\nabla_0 \dot{\phi}$, $\mathbf{M}$ and $\nabla_0 \mathbf{M}$, when considering the decomposition of the overall Helmholtz free energy in terms of a skeleton and a fluid potential; $\phi$, $\nabla_0 \phi$ represent the Lagrangean porosity $\phi = nJ$ and its gradient. Avoiding to consider the aforementioned decomposition, the Clausius–Duhem inequality will be written in terms of $\dot{\mathbf{\Lambda}}$, $\nabla_0 \dot{\mathbf{\Lambda}}$, $\dot{m}_f$, $\nabla_0 \dot{m}_f$, $\mathbf{M}$ and $\nabla_0 \mathbf{M}$.

According to the requirements of the second principle of thermodynamics a suitable representation formula for the constitutive relations of (generalized) stresses can be determined for a fluid which fills the whole pore space of a porous solid skeleton; these constitutive relations are summarized in the following formulas,

---

[3] We remind that an apparent kinematical parameter is that which guarantees the $a$th mass balance to be written in terms the density per unit volume of the mixture.

$$\frac{\partial e_{\mathrm{f}}}{\partial(1/\rho_{\mathrm{f}})} = -\left[p_{\mathrm{f}}^{\mathrm{c}} + \left(1 + \frac{1}{\mathrm{tr}\mathbf{I}}\right)\frac{\mathbf{c}_{\mathrm{f}}}{n\rho_{\mathrm{f}}} \cdot \nabla(n\rho_{\mathrm{f}})\right], \tag{26}$$

$$\frac{\partial e_{\mathrm{f}}}{\partial \nabla(n\rho_{\mathrm{f}})} = -\frac{n\mathbf{c}_{\mathrm{f}}}{(n\rho_{\mathrm{f}})^2}, \tag{27}$$

$$\mathrm{dev}\mathbf{T}_{\mathrm{f}}^{\mathrm{c}} = -\rho_{\mathrm{f}}\nabla(n\rho_{\mathrm{f}}) \otimes \frac{\partial e_{\mathrm{f}}}{\partial \nabla(n\rho_{\mathrm{f}})}, \tag{28}$$

where dev indicates the deviatoric part of a second order tensor. We refer to Casal (1972) and Seppecher (1987) for more details. We underline that $p_{\mathrm{f}}^{\mathrm{c}}$, $\mathrm{dev}\mathbf{T}_{\mathrm{f}}^{\mathrm{c}}$ and $\mathbf{c}_{\mathrm{f}}$ are the (generalized) stresses of the pure fluid; they are defined so as to guarantee Eq. (9) to hold true for the fluid constituent. In particular we remark that the driving idea in the identification of the different contributions to the (time) variation of $e_{\mathrm{f}}$ (see Eqs. (26)–(28)) is the same as that used in classical thermodynamics; the fluid pressure is the thermodynamic force which works on the divergence of velocity, the deviatoric stress tensor is the force which works on the deviatoric part of the velocity gradient, moreover the fluid hyper-stress is the intensive thermodynamic parameter working on the second gradient of the velocity (or gradient of the divergence of the velocity, in the considered case).

According to the previous remarks the hyper-stress vector $\mathbf{c}_{\mathrm{f}}$ depends on the gradient of the specific density of the fluid as well as on the gradient of porosity. We remark that in Eqs. (26) and (28) a special character $c$ has been introduced to underline that only the conservative part of the fluid pressure and deviatoric stresses is prescribed by these formulas; the dissipative contribution, if any, will be characterized by the Clausius–Duhem inequality. No special character labels the hyper-stress vector field $\mathbf{c}_{\mathrm{f}}$ since we aim to account only for purely conservative contributions working on the second gradient of the velocity.

This representation of the second gradient fluid-specific internal energy provides a proper description of surface tension effects and, in particular, of internal capillarity and co-capillarity phenomena (see e.g. Seppecher, 1987). Apparently, the derivative of the fluid internal energy with respect to the specific entropy equals the absolute temperature $T$, as in the classical first gradient theory.

In the following, two different deductions of the constitutive theory will be developed according to the representation formula of the overall Helmholtz free energy which we shall account for; in particular we shall distinguish the case in which no additional hypotheses will be assumed for this functional from the case in which it could be split into a solid and a fluid potential:

$$\Psi = \Psi_{\mathrm{s}} + m_{\mathrm{f}}\psi_{\mathrm{f}}|_{(n,\nabla n)=\mathrm{const}}. \tag{29}$$

In this formula $\Psi_{\mathrm{s}}$ is the skeleton Lagrangean density of free energy, whilst $\psi_{\mathrm{f}} = e_{\mathrm{f}} - Ts_{\mathrm{f}}$ is the specific free energy of the fluid in terms of the internal energy, entropy and absolute temperature.

Let us now introduce the first and the second principle of thermodynamics; starting in particular from an Eulerian form of the two principles we deduce the corresponding Lagrangean pull back in the reference configuration of the solid skeleton.

### 4.1. The first principle

In its general form the first principle of thermodynamics for a porous continuum accounts for the variation of the solid and fluid internal energies following the motion of solid and fluid particles, respectively:

$$\frac{\mathrm{d}^{\mathrm{s}}}{\mathrm{d}t}\int_{\mathcal{D}}\rho_{\mathrm{s}}(1-n)e_{\mathrm{s}}\,\mathrm{d}v + \frac{\mathrm{d}^{\mathrm{f}}}{\mathrm{d}t}\int_{\mathcal{D}}\rho_{\mathrm{f}}ne_{\mathrm{f}}\,\mathrm{d}v = \mathcal{W}^{\mathrm{int}} + \mathring{\mathcal{Q}}, \quad \mathring{\mathcal{Q}} := -\int_{\partial\mathcal{D}}\mathbf{q}\cdot\mathbf{n}\,\mathrm{d}s, \tag{30}$$

where $\mathring{\mathcal{Q}}$ is the rate of heat externally supplied and $\mathbf{q}$ the heat flow vector; moreover $\mathrm{d}^a/\mathrm{d}t$ ($a = s,f$) represents the material derivative following the motion of the $a$th constituent. Let us introduce the overall internal energy (24), $e$ and its Lagrangean pull back in the reference configuration of the solid skeleton $E = Je$; Eq. (30) reads now as

$$\frac{\mathrm{d}}{\mathrm{d}t}\int_{\mathcal{D}_0}E\,\mathrm{d}v_0 = \mathcal{W}_{\mathcal{L}}^{\mathrm{int}} - \int_{\mathcal{D}_0}\mathrm{div}_0(e_{\mathrm{f}}\mathbf{M} + \mathbf{Q})\,\mathrm{d}v_0, \quad \mathbf{Q} := J\mathbf{F}^{-1}\mathbf{q}, \tag{31}$$

where d/dt indicates the material time derivative following the motion of the solid skeleton. According to Eq. (23), it can be written in the local form

$$\frac{dE}{dt} = [S - C^{-1}((\nabla_0 C)\gamma)C^{-1}] \cdot \frac{d\Delta}{dt} + C^{-1} \otimes \gamma \cdot \nabla_0 \frac{d\Delta}{dt} - \text{div}_0\left[\left(e_f + \frac{p_f}{\rho_f} + \frac{1}{m_f}\text{div}_0 n\gamma_f\right)M\right]$$
$$+ \text{div}_0\left[\left(\frac{1}{\text{tr}\mathbf{I}}\frac{n\gamma_f}{J} \cdot \nabla_0 \frac{1}{n\rho_f}\right)M + \frac{\gamma_f}{J\rho_f}\text{div}_0 M - Q\right] + \text{div}_0\left[\text{dev}\left(\frac{1}{\rho_f}S_f^\top + \left(n\gamma_f \otimes \nabla_0 \frac{1}{n\rho_f}\right)\right)\frac{M}{J}\right]. \quad (32)$$

In the following we use the overall Lagrangean Helmholtz free energy $\Psi = E - TS$ instead of the overall internal energy $E$; $S$ is the overall Lagrangean density of entropy.

This very general form obviously simplifies when considering only a linear problem, which will be our choice in the next sections:

$$\frac{dE}{dt} = \sigma \cdot \frac{d\varepsilon}{dt} + \mathbf{I} \otimes \gamma \cdot \nabla_0 \frac{d\varepsilon}{dt} + \text{div}_0\left[\frac{\gamma_f}{\rho_f^0}\text{div}_0 M - \left(e_f + \frac{v}{\rho_f^0}\right)M + \text{dev}S_f^\top \frac{M}{\rho_f^0} - Q\right]. \quad (33)$$

In particular, in this last equation we assumed for the sake of simplicity the initial configuration of the porous medium to be macroscopically homogeneous. In Eq. (33) $\sigma$ and $\varepsilon$ represent the linearized stress and strain fields; with an abuse of notation $p_f$, $\gamma_f$ and $\gamma$ indicate the linearized (generalized) stresses of the fluid and the hyper-stress of the skeleton.

### 4.2. The second principle

We now do the same for the integral form of the second principle of thermodynamics in the current configuration of the porous medium:

$$\frac{d^s}{dt}\int_\mathcal{D} \rho_s(1-n)s_s dv + \frac{d^f}{dt}\int_\mathcal{D} \rho_f n s_f dv \geq -\int_{\partial\mathcal{D}} \frac{\mathbf{q} \cdot \mathbf{n}}{T} ds. \quad (34)$$

Following the same procedure used in order to determine the local form of Eq. (30) pulled back in the reference configuration of the skeleton, we introduce the representation of the overall Lagrangean density of entropy $S = J(\rho_s(1-n)s_s + \rho_f n s_f) = S_s + m_f s_f$ and therefore determine the following local inequality:

$$\frac{dE}{dt} - S\frac{dT}{dt} - \frac{d\Psi}{dt} \geq -T\text{div}_0\left(s_f M - \frac{Q}{T}\right). \quad (35)$$

Substituting Eq. (32) into inequality (35) we derive the Clausius–Duhem inequality valid for a second gradient porous continuum

$$\Phi = \Phi_s + \Phi_f + \Phi_{th} \geq 0, \quad (36)$$

where the solid dissipation, $\Phi_s$, the fluid dissipation, $\Phi_f$, and the thermal dissipation, $\Phi_{th}$, have the form

$$\Phi_s = \left[S - C^{-1}((\nabla_0 C)\gamma)C^{-1} - [C^{-1}\nabla_0(n\rho_f)] \otimes \frac{\gamma_f}{\rho_f}\right] \cdot \frac{d\Delta}{dt} + C^{-1} \otimes \gamma \cdot \frac{d\nabla_0\Delta}{dt} - \frac{\gamma_f}{J\rho_f} \cdot \frac{d}{dt}\nabla_0 m_f$$
$$+ \left[g_f - \left(1 + \frac{1}{\text{tr}\mathbf{I}}\right)n\gamma_f \cdot \nabla_0\left(\frac{1}{m_f}\right) + \frac{1}{\text{tr}\mathbf{I}}\frac{\gamma_f}{\rho_f} \cdot \nabla_0(J^{-1})\right]\frac{dm_f}{dt} - S\frac{dT}{dt} - \frac{d\Psi}{dt}, \quad (37)$$

$$\Phi_f = -\nabla_0\left[e_f + \frac{p_f^c + p_f^d}{\rho_f} + \frac{1}{m_f}\text{div}_0(n\gamma_f) + \frac{1}{\text{tr}\mathbf{I}}\frac{\gamma_f}{J} \cdot \left(\frac{1}{\rho_f^2}\nabla_0\rho_f + \frac{\nabla_0 n}{n\rho_f}\right)\right] \cdot M$$
$$+ \text{div}_0\left(\frac{1}{J\rho_f}\text{dev}S_f^d\right) \cdot M + \left(\frac{1}{J\rho_f}\text{dev}S_f^d\right) \cdot \nabla_0 M - \frac{p_f^d}{\rho_f}\text{div}_0 M + T\nabla_0 s_f \cdot M, \quad (38)$$

$$\Phi_{th} = -\frac{Q}{T} \cdot \nabla_0 T. \quad (39)$$

If the decomposition (29) holds Eq. (37) must be replaced by

$$\Phi_s = \left[ S - C^{-1}((\nabla_0 C)\gamma)C^{-1} - [C^{-1}\nabla_0(n\rho_f)] \otimes \frac{\gamma_f}{\rho_f} \right] \cdot \frac{d\Delta}{dt} + C^{-1} \otimes \gamma \cdot \frac{d\nabla_0 \Delta}{dt} - \gamma_f \cdot \frac{d}{dt}\nabla_0 \phi$$
$$+ \left( p_f^c + \frac{1}{\text{tr}\mathbf{I}} \frac{\gamma_f}{J} \cdot \frac{1}{n\rho_f} \nabla_0(n\rho_f) - \frac{\gamma_f}{J} \cdot \phi \nabla_0 \frac{1}{\phi} \right) \frac{d\phi}{dt} - S_s \frac{dT}{dt} - \frac{d\Psi_s}{dt}. \quad (40)$$

Deduction of Eq. (40) is presented in Appendix A.

In the fluid dissipation, additional terms associated with dissipative contributions to the fluid-stress tensor were introduced; these terms identify a kind of Brinkman extension of the Darcy law (see Brinkman, 1947).

Owing to the different nature of the dissipations previously identified we assume the decoupling hypothesis to hold true, substituting the unique inequality (36) by three separate inequalities:

$$\Phi_s \geqslant 0, \quad \Phi_f \geqslant 0, \quad \Phi_{th} \geqslant 0. \quad (41)$$

Solid dissipation states that, during isothermal processes, the strain work rate which can be stored by the overall porous material in the form of free energy, is related not only to first and second gradient of the solid displacement but also to the Lagrangean fluid mass density and its gradient. This means that the mechanical work of the overall body depends both on the solid strain (and the solid strain gradient) and the fluid mass density (and the fluid mass gradient), which indeed is a well-known result in classical poromechanics (see e.g. Coussy, 2004). Fluid dissipation also deserves attention concerning in particular the pressure like correcting terms, due to second gradient, and the non-vanishing contribution related to the presence of deviatoric stresses.

From now on the solid dissipation will be required to vanish ($\Phi_s = 0$), which means that only conservative internal actions of the skeleton will be modeled. This assumption means that the overall Helmholtz free energy as well as the Helmholtz free energy of the solid skeleton of the form (29) exhibit the dependancies

$$\Psi = \hat{\Psi}(\Delta, \nabla_0 \Delta, m_f, \nabla_0 m_f, T), \qquad \Psi_s = \hat{\Psi}_s(\Delta, \nabla_0 \Delta, \phi, \nabla_0 \phi, T); \quad (42)$$

in other words one neglects the dependence of $\Psi$ and $\Psi_s$ on additional internal state variables $\chi_J (J = 1, \ldots, N)$ associated to irreversible processes occurring inside the skeleton.

It is worth noticing that the kinematic fields appearing on the right-hand side of Eq. (42) guarantee the macroscopic potentials $\Psi$ and $\Psi_s$ to be invariant under changes of observer: as a matter of facts none of them is affected by the rule of change of the spatial gradient of the placement map ($F^* = QF$, $F$ and $F^*$ being the representation of the same tensor field in two different frames). This property is only related to the fact that strain gradients, appearing in the Helmholtz free energy, are Lagrangean gradients and therefore are intrinsically objective fields; this is not the case when the second gradient model is formulated in the current configuration of the mixture, as it is for the case of second gradient fluids (see e.g. Cahn and Hilliard, 1959). This remark is a crucial step for further developments: as a matter of facts according to the objective dependence of $\hat{\Psi}$ on ($\Delta$, $\nabla_0 \Delta$, $m_f$, $\nabla_0 m_f$) and $\hat{\Psi}_s$ on ($\Delta$, $\nabla_0 \Delta$, $\phi$, $\nabla_0 \phi$) we can introduce in the constitutive model not only purely second gradient corrections, having the same structure as those ones considered for modeling wetting phenomena in fluids (see e.g. de Gennes, 1985) but also mixed terms with respect to first and second gradient kinematic parameters.

Regarding the dissipation of the fluid we notice that Eq. (38) generalizes the standard dissipation formula obtained for a first gradient poromechanical model: such a dissipation involves not only the filtration vector $M$ but also the gradient of $M$ which has been split into its spherical and deviatoric part: $\text{div}_0 M$ and $\text{dev}\nabla_0 M$, respectively. These two terms are possibly associated to the case of a saturating fluid which exhibits a viscous behaviour at the micro level. The macroscopic diffusion is described in this case via a kind of Brinkman law.

In the following two sections we shall introduce the second gradient constitutive characterization of the conservative stresses, for the overall and the pure fluid material, as well as the dissipative ones, which will be only those of the fluid. For what concern dissipative phenomena, following the classical procedure of Coleman and Noll (1963) we shall establish sufficient constitutive assumptions which guarantee the dissipation inequality to hold true; these last will turn to be necessary and sufficient for satisfying the second principle of thermodynamics in the linearized problem.

## 5. Constitutive law

Let us now discuss how to determine the constitutive relations for the porous material starting from the expression of the overall Helmholtz free energy or from the Helmholtz free energy of the skeleton given in Eq. (42). Apparently the material time derivatives of $\Psi$ and $\Psi_s$ read

$$\frac{d\Psi}{dt} = \frac{\partial\hat{\Psi}}{\partial\mathbf{\Delta}} \cdot \frac{d\mathbf{\Delta}}{dt} + \frac{\partial\hat{\Psi}}{\partial\nabla_0\mathbf{\Delta}} \cdot \frac{d}{dt}\nabla_0\mathbf{\Delta} + \frac{\partial\hat{\Psi}}{\partial m_f}\frac{dm_f}{dt} + \frac{\partial\hat{\Psi}}{\partial\nabla_0 m_f}\frac{d}{dt}\nabla_0 m_f + \frac{\partial\hat{\Psi}}{\partial T}\frac{dT}{dt}, \tag{43}$$

$$\frac{d\Psi_s}{dt} = \frac{\partial\hat{\Psi}_s}{\partial\mathbf{\Delta}} \cdot \frac{d\mathbf{\Delta}}{dt} + \frac{\partial\hat{\Psi}_s}{\partial\nabla_0\mathbf{\Delta}} \cdot \frac{d}{dt}\nabla_0\mathbf{\Delta} + \frac{\partial\hat{\Psi}_s}{\partial\phi}\frac{d\phi}{dt} + \frac{\partial\hat{\Psi}_s}{\partial\nabla_0\phi}\frac{d}{dt}\nabla_0\phi + \frac{\partial\hat{\Psi}_s}{\partial T}\frac{dT}{dt}, \tag{44}$$

which implies the following constitutive relations for the internal forces $\mathbb{S}$, $\gamma$, $p_f^c$ and $\gamma_f$, requiring the solid dissipation to vanish:

$$\frac{\partial\hat{\Psi}}{\partial\mathbf{\Delta}} = \frac{\partial\hat{\Psi}_s}{\partial\mathbf{\Delta}} = \mathbb{S} - \mathbf{C}^{-1}((\nabla_0\mathbf{C})\gamma)\mathbf{C}^{-1} - (\mathbf{C}^{-1}\nabla_0\rho_f) \otimes \frac{\gamma_f\phi}{J\rho_f}, \tag{45}$$

$$\frac{\partial\hat{\Psi}}{\partial\nabla_0\mathbf{\Delta}} = \frac{\partial\hat{\Psi}_s}{\partial\nabla_0\mathbf{\Delta}} = \mathbf{C}^{-1} \otimes \gamma, \tag{46}$$

$$\frac{\partial\hat{\Psi}}{\partial m_f} = e_f + \frac{p_f^c}{\rho_f} - \left(1 + \frac{1}{\mathrm{tr}\mathbf{I}}\right)\frac{n\gamma_f}{J} \cdot \nabla_0\left(\frac{1}{n\rho_f}\right) - \frac{\gamma_f}{J\rho_f} \cdot \nabla_0 J^{-1}, \quad \frac{\partial\hat{\Psi}_s}{\partial\nabla_0\phi} = -\gamma_f, \tag{47}$$

$$\frac{\partial\hat{\Psi}_s}{\partial\phi} = p_f^c + \frac{\gamma_f}{J} \cdot \left(\frac{1}{\mathrm{tr}\mathbf{I}}\frac{1}{\rho_f}\nabla_0\rho_f - \phi\nabla_0\frac{1}{\phi}\right), \quad \frac{\partial\hat{\Psi}}{\partial\nabla_0 m_f} = -\frac{\gamma_f}{J\rho_f}, \tag{48}$$

$$\frac{\partial\hat{\Psi}}{\partial T} = -S, \quad \frac{\partial\hat{\Psi}_s}{\partial T} = -S_s. \tag{49}$$

As we have already noticed in the previous section the dependence of $\hat{\Psi}$ and $\hat{\Psi}_s$ on the state parameters $(\mathbf{\Delta}, \nabla_0\mathbf{\Delta}, m_f, \nabla_0 m_f, T)$ and $(\mathbf{\Delta}, \nabla_0\mathbf{\Delta}, \phi, \nabla_0\phi, T)$, respectively, guarantees the energy functionals to be frame indifferent, i.e. invariant under a change of observer; this means that every internal force may depend on the whole set of state parameters.

In the linearized case, i.e. when considering small deformation about a stress-free reference configuration of the solid skeleton and assuming isothermal processes, the constitutive laws reduce to

$$\frac{\partial\hat{\Psi}}{\partial\boldsymbol{\varepsilon}} = \frac{\partial\hat{\Psi}_s}{\partial\boldsymbol{\varepsilon}} = \boldsymbol{\sigma}, \quad \frac{\partial\hat{\Psi}}{\partial\nabla\boldsymbol{\varepsilon}} = \frac{\partial\hat{\Psi}_s}{\partial\nabla\boldsymbol{\varepsilon}} = \mathbf{I} \otimes \gamma, \tag{50}$$

$$\frac{\partial\hat{\Psi}}{\partial m_f}\rho_f^0 = \frac{\partial\hat{\Psi}_s}{\partial\phi} = p_f^c, \quad \frac{\partial\hat{\Psi}}{\partial\nabla m_f}\rho_f^0 = \frac{\partial\hat{\Psi}_s}{\partial\nabla\phi} = -\gamma_f, \tag{51}$$

where $\boldsymbol{\sigma}$ indicates the linearized Piola–Kirchhoff stress and $\boldsymbol{\varepsilon}$ the small strain; moreover, $p_f^c$ and $\gamma_f$ indicate with an abuse of notation the linearized stresses and hyper-stresses. In this case the Eulerian and Lagrangean gradients coincide and consequently no subscript will label nabla operators in the following. According to Eq. (50)$_2$, $\hat{\Psi}$ and $\hat{\Psi}_s$ do not depend on the complete $\nabla\boldsymbol{\varepsilon}$ but only on $\nabla\mathrm{tr}\boldsymbol{\varepsilon}$, which means that $\gamma = \partial\hat{\Psi}/\partial\nabla\mathrm{tr}\boldsymbol{\varepsilon} = \partial\hat{\Psi}_s/\partial\nabla\mathrm{tr}\boldsymbol{\varepsilon}$.

Let us now separately discuss constitutive relations which can be obtained by starting from the overall Helmholtz free energy or from the corresponding potential relative to the skeleton only. We decide, in particular, to consider the following formulas for the addressed linearized problem, where for the sake of simplicity we just consider isotropic second gradient effects:

$$\Psi = \frac{1}{2}\left\{\mathsf{C}_s[\boldsymbol{\varepsilon}]\cdot\boldsymbol{\varepsilon} + M(\mathsf{B}\cdot\boldsymbol{\varepsilon})^2\right\} + \frac{1}{2}M\left(\frac{\Delta m_f}{\rho_f^0}\right)^2 - M(\mathsf{B}\cdot\boldsymbol{\varepsilon})\frac{\Delta m_f}{\rho_f^0} + \left(\mathbb{T}_{ss}\nabla\mathrm{tr}\boldsymbol{\varepsilon} + \mathbb{T}_{sf}\frac{\nabla\Delta m_f}{\rho_f^0}\right)\cdot\boldsymbol{\varepsilon}$$

$$+ \left(\boldsymbol{\gamma}_{ss}\cdot\nabla\mathrm{tr}\boldsymbol{\varepsilon} + \boldsymbol{\gamma}_{sf}\cdot\frac{\nabla\Delta m_f}{\rho_f^0}\right)\frac{\Delta m_f}{\rho_f^0} + \frac{1}{2}\left\{\mathbb{K}_{ss}||\nabla\mathrm{tr}\boldsymbol{\varepsilon}||^2 + \mathbb{M}\mathsf{k}_{sf}^2||\nabla\mathrm{tr}\boldsymbol{\varepsilon}||^2\right\} + \frac{1}{2}\mathbb{M}\frac{||\nabla\Delta m_f||^2}{\rho_f^0}$$

$$+ \mathbb{M}\left(\mathsf{k}_{sf}\nabla\mathrm{tr}\boldsymbol{\varepsilon}\cdot\frac{\nabla\Delta m_f}{\rho_f^0}\right), \tag{52}$$

$$\Psi_s^* = \frac{1}{2}\mathsf{C}_s[\boldsymbol{\varepsilon}]\cdot\boldsymbol{\varepsilon} - \mathsf{B}\cdot\boldsymbol{\varepsilon}\Delta p_f^c - \frac{1}{2N}\Delta p_f^{c2} + (\mathbb{L}_{ss}\nabla\mathrm{tr}\boldsymbol{\varepsilon} + \mathbb{L}_{sf}\nabla\Delta\phi)\cdot\boldsymbol{\varepsilon}$$

$$- (\boldsymbol{\ell}_{fs}\cdot\nabla\mathrm{tr}\boldsymbol{\varepsilon} + \boldsymbol{\ell}_{ff}\cdot\nabla\Delta\phi)\Delta p_f^c + \frac{1}{2}\boldsymbol{\Lambda}_{ss}||\nabla\mathrm{tr}\boldsymbol{\varepsilon}||^2 - \boldsymbol{\Lambda}_{sf}\nabla\Delta\phi\cdot\nabla\mathrm{tr}\boldsymbol{\varepsilon} + \frac{1}{2}\boldsymbol{\Lambda}_{ff}||\nabla\Delta\phi||^2. \tag{53}$$

Eqs. (52) and (53) define the overall Helmholtz free energy and a Legendre transformation of that of the specific solid in the envisaged frame of linearized theory:

$$\Psi_s^* = \Psi_s - p_f^c\phi; \tag{54}$$

apparently from this last formula it follows that $\partial\hat{\Psi}_s^*/\partial p_f^c = -\phi$ replaces Eq. $(51)_1$ in the constitutive relations deduced from the skeleton potential; the new set of state parameters (extensive variables) is then $\boldsymbol{\varepsilon}$, $\nabla\mathrm{tr}\,\boldsymbol{\varepsilon}$, $p_f^c$, $\nabla\phi$.

As we already noticed, frame indifference of the overall and skeleton potentials implies that every intensive variable (i.e. every thermodynamic force) can depend on the complete set of the conjugate extensive parameters; accordingly the following formulation of linearized constitutive relations can be stated:

(i) when prescribing the overall Helmholtz free energy – see Eq. (52)

$$\boldsymbol{\sigma} = (\mathsf{C}_s - M\mathsf{B}\otimes\mathsf{B})[\boldsymbol{\varepsilon}] - M\mathsf{B}\frac{\Delta m_f}{\rho_f^0} + \mathbb{T}_{ss}\nabla\mathrm{tr}\boldsymbol{\varepsilon} + \mathbb{T}_{sf}\frac{\nabla\Delta m_f}{\rho_f^0}, \tag{55}$$

$$\Delta p_f^c = M\frac{\Delta m_f}{\rho_f^0} - M\mathsf{B}\cdot\boldsymbol{\varepsilon} + \boldsymbol{\gamma}_{ss}\cdot\nabla\mathrm{tr}\boldsymbol{\varepsilon} + \boldsymbol{\gamma}_{sf}\cdot\frac{\nabla\Delta m_f}{\rho_f^0}, \tag{56}$$

$$\boldsymbol{\gamma} = \mathbb{T}_{ss}^\top\boldsymbol{\varepsilon} + \boldsymbol{\gamma}_{ss}\frac{\Delta m_f}{\rho_f^0} + (\mathbb{K}_{ss} + \mathbb{M}\mathsf{k}_{sf}^2)\nabla\mathrm{tr}\boldsymbol{\varepsilon} + \mathbb{M}\mathsf{k}_{sf}\frac{\nabla\Delta m_f}{\rho_f^0}, \tag{57}$$

$$\boldsymbol{\gamma}_f = \mathbb{T}_{sf}^\top\boldsymbol{\varepsilon} + \boldsymbol{\gamma}_{sf}\frac{\Delta m_f}{\rho_f^0} + \mathbb{M}\nabla\mathrm{tr}\boldsymbol{\varepsilon} + \mathbb{M}\mathsf{k}_{sf}\frac{\nabla\Delta m_f}{\rho_f^0} \tag{58}$$

(ii) when assuming the Helmholtz free energy of the skeleton – see Eq. (53)

$$\boldsymbol{\sigma} = \mathsf{C}_s[\boldsymbol{\varepsilon}] - \mathsf{B}\Delta p_f^c + \mathbb{L}_{ss}\nabla\mathrm{tr}\boldsymbol{\varepsilon} + \mathbb{L}_{sf}\nabla\Delta\phi, \tag{59}$$

$$\Delta\phi = \mathsf{B}\cdot\boldsymbol{\varepsilon} + \frac{1}{N}\Delta p_f^c + \boldsymbol{\ell}_{fs}\cdot\nabla\mathrm{tr}\boldsymbol{\varepsilon} + \boldsymbol{\ell}_{ff}\cdot\nabla\Delta\phi, \tag{60}$$

$$\boldsymbol{\gamma} = \mathbb{L}_{ss}^\top\boldsymbol{\varepsilon} - \boldsymbol{\ell}_{fs}\Delta p_f^c + \boldsymbol{\Lambda}_{ss}\nabla\mathrm{tr}\boldsymbol{\varepsilon} - \boldsymbol{\Lambda}_{sf}\nabla\Delta\phi, \tag{61}$$

$$\boldsymbol{\gamma}_f = -\mathbb{L}_{sf}^\top\boldsymbol{\varepsilon} + \boldsymbol{\ell}_{ff}\Delta\phi + \boldsymbol{\Lambda}_{sf}\nabla\mathrm{tr}\boldsymbol{\varepsilon} - \boldsymbol{\Lambda}_{ff}\nabla\Delta\phi. \tag{62}$$

According to classical (poro-) mechanics $\mathsf{C}_s$ is the elastic stiffness tensor of the solid skeleton, $\mathsf{B}$ the Biot tangent tensor and $1/N$ the inverse Biot tangent modulus; $M$ is related to $N$ according to the following condition (see e.g. Coussy, 2004): $1/M = 1/N + \phi_0/K_f$, where $\phi_0$ is the reference Lagrangean porosity and $K_f$ the bulk modulus of the fluid. Moreover $\mathbb{K}_{ss}$, $\mathsf{k}_{sf}$ and $\mathbb{M}$, as well as $\boldsymbol{\Lambda}_{ss}$, $\boldsymbol{\Lambda}_{sf}$, $\boldsymbol{\Lambda}_{ff}$, describe purely second gradient poroelastic properties, which relate second gradient intensive variables ($\boldsymbol{\gamma}$ and $\boldsymbol{\gamma}_f$) to the gradient of strain and Lagrangean fluid mass content, or porosity, in case of Eqs. (59)–(62). Finally, $\mathbb{T}_{ss}$, $\mathbb{T}_{sf}$, $\boldsymbol{\gamma}_{ss}$ and $\boldsymbol{\gamma}_{sf}$, as well as $\boldsymbol{\ell}_{fs}$, $\boldsymbol{\ell}_{ff}$, $\mathbb{L}_{ss}$, $\mathbb{L}_{sf}$ describe the poroelastic properties which couple first gradient strain measures to second gradient intensive variables and, vice-versa, gradients of strain and fluid mass (porosity) to first gradient intensive variables.

It is worth noticing that the mixed terms involved in Eqs. (55)–(58) as well as (59)–(62) are capable incorporating non-local strain effects when modeling stress constitutive laws and conversely pure strain effects when modeling the constitutive relation of hyper-stress. These terms are capable of accounting for the effect of non-local strain and fluid-mass (porosity) changes on the value of local stresses; in other words they consider how strain energy density associated to a given RVE depends on the energy of the neighbourings. The macroscopic distribution of stresses in the vicinity of the RVE guarantees not only local equilibrium of the volume and the neighbourings but also surface equilibrium at the boundary of contiguous RVEs. Even when starting from a macroscopically homogeneous initial configuration, fluid mass (porosity) and volume-change gradients can be encountered because, for instance, of non-homogeneous filtration; consequently local stresses arise in the occurrence of a kind of counter drift-like phenomenon.

Conversely local porosity (or local pore pressure) and local volume changes can affect both skeleton and pore hyper-stresses ($\gamma$ and $\gamma_f$, respectively).

## 6. The extended Darcy law

In this section, we shall deduce from the fluid dissipation inequality (38) a suitable extension of the classical Darcy law; in particular corrective terms of the fluid pressure, associated to the divergence of the pore hyper-stress, as well as additional diffusion terms, associated with second derivatives of the filtration vector $\mathbf{M}$, are going to be identified in the evolution equation for the motion of the fluid with respect to the solid.

Since the filtration vector $\mathbf{M}$ does not equal the rate of change of any state parameter, then, according to the classical procedure stated by Coleman and Noll (1963), a sufficient condition for the second principle of thermodynamics to hold is to regard the fluid dissipation as a quadratic form of $\mathbf{M}$, $\mathrm{div}_0\mathbf{M}$ and $\mathrm{dev}\nabla_0\mathbf{M}$; this yields the following conditions

$$\frac{p_f^d}{\rho_f} = -\alpha \mathrm{div}_0\mathbf{M}; \quad \frac{1}{J\rho_f}\mathrm{dev}\mathbb{S}_f^d = \mathbb{A}\mathrm{dev}\nabla_0\mathbf{M}, \tag{63}$$

$$-\frac{1}{\rho_f}\nabla_0\left[p_f^c + \frac{1}{\phi}\mathrm{div}_0(n\gamma_f) + \frac{1}{\mathrm{tr}\mathbf{I}}\frac{\gamma_f}{J} \cdot \frac{1}{n\rho_f}\nabla_0(n\rho_f)\right] + [(\nabla_0\mathbf{F}^{-\top})\nabla_0\rho_f]^{\top}\frac{\mathbf{F}\gamma_f}{J\rho_f^2} + \nabla_0\nabla_0\rho_f\frac{\gamma_f}{J\rho_f^2}$$

$$-\left[\frac{1}{\phi}\mathrm{div}_0(n\gamma_f) - \frac{\gamma_f}{m_f} \cdot \nabla_0(n\rho_f)\right]\nabla_0\frac{1}{\rho_f} + \nabla_0(\alpha\mathrm{div}_0\mathbf{M}) + \mathrm{div}_0(\mathbb{A}\mathrm{dev}\nabla_0\mathbf{M}) = \mathbf{A}\mathbf{M}. \tag{64}$$

Accordingly, Eq. (38) takes now the form

$$\Phi_f = \mathbf{A}\mathbf{M} \cdot \mathbf{M} + \mathbb{A}\mathrm{dev}\nabla_0\mathbf{M} \cdot \mathrm{dev}\nabla_0\mathbf{M} + \alpha(\mathrm{div}_0\mathbf{M})^2, \tag{65}$$

where $\mathbf{A}$ is a second order tensor related to the permeability $\mathbf{K} = -J\rho_f^2\mathbf{F}\mathbf{A}^{-1}$, $\mathbb{A}$ is a fourth order tensor defined over the space of second order deviatoric tensors and $\alpha$ is a scalar parameter.

In the linearized case, i.e. when considering small deformations about a stress-free reference configuration of the solid skeleton, the Darcy law is given by

$$-\frac{1}{\rho_f}\nabla_0(p_f^c + \mathrm{div}_0\gamma_f) = \mathbf{A}\mathbf{M} - \nabla_0(\alpha\mathrm{div}_0\mathbf{M}) - \mathrm{div}_0(\mathbb{A}\mathrm{dev}\nabla_0\mathbf{M}); \tag{66}$$

in the isotropic case the second order tensor $\mathbf{A}$ is univocally determined by a scalar parameter, such that $\mathbf{A} = a\mathbf{I}$, moreover the fourth order tensor $\mathbb{A}$ acts on deviatoric second order tensors according to the following condition: $\mathbb{A}\mathrm{dev}\nabla_0\mathbf{M} = a_1\mathrm{sym}(\mathrm{dev}\nabla_0\mathbf{M}) + a_2\mathrm{skw}(\mathrm{dev}\nabla_0\mathbf{M}) - a_1$ and $a_2$ being suitable scalar parameters.

As we already noticed the additional friction terms involved in Eq. (64) describe dissipative phenomena which are not accounted for in the classical Darcy law. Indeed they are dissipative terms associated to second derivatives of the fluid mass vector, which represents the Lagrangean pull back of the velocity in the reference configuration of the skeleton; this additional dissipation implies the equation governing the behaviour of the fluid to be of Brinkman kind (see Brinkman, 1947). As it is well known this is the extension of the Darcy law when the size of the obstacle to the motion of the fluid is sufficiently small with respect to the characteristic size of the inter-obstacle distance (a rational up-scaling of the Navier Stokes equation towards the Darcy and the Brinkman law has been stated in Allaire (1997) for the case of a microscopically incompressible fluid).

Considering a suitable discretization of a prototype one-dimensional problem the introduced corrective terms identify a dependence of the pressure gradient at $X$ on the relative velocity of the fluid with respect to the solid in $X$, $X + u$ and $X - u$, $u$ being the spatial discretization step. With a naive microscopic interpretation we can say that a viscous boundary layer at the solid–fluid interphase exists at the pore scale; this is sufficiently large to overwhelm from one pore to its neighbour.

## 7. The consolidation problem: an extension of the classical Terzaghi equations

In this section, we shall produce a rational extension in the frame of second gradient poromechanics of the classical results on consolidation theory, due to Terzaghi (see e.g. de Boer, 1996; Coussy, 2004). The problem is defined over a one-dimensional space interval (the porous medium) and the time axis; the goal is to describe how the spatial profiles of the fluid pressure change in time when the fluid flows out of the porous material because of the consolidation pressure ($\bar{p}$) applied on the boundary of the porous material. This boundary condition is described by means of a Heaviside function, concentrated at time $t = 0$. The fluid flow is modeled by the Darcy law, therefore no inertia effect is considered.

In this section, we restrict our attention to the linearized problem, i.e. the one where only small deformations around a stress-free reference configuration of the solid skeleton are taken into account. We regard the linearized classical Biot poromechanical model as a milestone; we consider the formulation of the model as presented in Coussy (2004), in particular as regards the initial conditions; we then compare the solution of the consolidation problem obtained in this frame with that relative to the introduced second gradient model. As it is well known, the classical Terzaghi solution does not provide a refined description of the fluid pressure behaviour close to the drained surface of the porous medium; as a matter of fact, no boundary layer effect can be envisaged by the classical model in the occurrence of the fluid flowing out of the body. As we shall see the proposed second gradient poromechanical model is capable of capturing a progressively decreasing value of the fluid pressure at the external surface starting from a non-vanishing initial value.

### 7.1. The classical Terzaghi consolidation problem

The perturbation of the pressure profile, with respect to its initial value, is determined in the standard framework as the solutions of the diffusive equation

$$\frac{\partial \Delta p}{\partial t} = c \nabla^2 (\Delta p) \tag{67}$$

endowed with the following initial and boundary conditions:

$$\Delta p(x, 0) + M b \epsilon(x, 0) = 0, \quad \forall x \in (0, l); \quad \Delta p(0, t) = 0, \ \mathbf{M}(l, t) = 0, \quad \forall t \in (0, \infty) \tag{68}$$

where $c$ denotes the diffusion coefficient, $c := kM(K + \frac{4}{3}\mu)/(K_u + \frac{4}{3}\mu)$; $b$ is the spherical component of Biot's tangent tensor, $\frac{1}{M} := \frac{1}{N} + \frac{\phi_0}{K_f}$ is a function of Biot's tangent modulus ($N$), of the fluid bulk modulus ($K_f$) and of the initial Lagrangean porosity ($\phi(t = 0) = \phi_0$), $k$ is the permeability coefficient, whilst $K$ and $K_u$ are the bulk modulus and the undrained bulk modulus of the skeleton.

The initial condition expresses that a kind of instantaneous equilibrium configuration is attained by the solid skeleton after the pressure $\bar{p}$ has been exerted on its free boundary; accordingly, no variation of the fluid content can be appreciated at this time; this statement is formalized by assuming that

$$\frac{m_f(t = 0^+) - m_f^0}{\rho_f^0} = 0, \quad m_f = \phi \rho_f, \tag{69}$$

or in other words, the instantaneous response of the porous material to the external loading is undrained. Bearing in mind that no bulk and inertia forces are considered in the model, the initial value of the fluid pressure is determined by considering the first integral associated to the balance equations of the solid, the corresponding boundary conditions, and the constitutive relation

$$\sigma' = 0, \quad \forall x \in (0, l) \tag{70}$$

$$\mathbf{t} = \sigma\mathbf{n} = [(1-n)\sigma_s + n\sigma_f]\mathbf{n} = -\bar{p}\mathbf{n}, \quad \text{in } x = 0, \ x = l \tag{71}$$

$$\sigma = (2\mu + \lambda)\epsilon - b\Delta p, \quad \text{in } \forall x \in (0, l); \tag{72}$$

which imply

$$\left(\frac{1}{M} + \frac{b^2}{2\mu + \lambda}\right)\Delta p(x, 0) - \frac{b\bar{p}}{2\mu + \lambda} = 0. \tag{73}$$

In Eqs. (71) and (72) the Eulerian porosity $n$ coincides with the Lagrangean porosity $\phi$ since a linearized problem is considered.

Boundary condition in $x = 0$ should describe drainage of the free surface of the porous material when $t \in (0^+, +\infty)$; this assumption requires that the variation of the traction exerted on the fluid vanishes

$$\mathbf{t}_f = n\sigma_f \mathbf{n} \Rightarrow \phi p = \phi_0 p_0. \tag{74}$$

Terzaghi's consolidation stays in the frame of a linearized model; therefore, Eq. (74) can be written in the following form

$$\phi_0 \Delta p + p_0 \Delta \phi = 0 \Rightarrow \Delta p + \overline{M} b \epsilon = 0, \quad \overline{M}^{-1} := \left(\frac{\phi_0}{p_0} + \frac{1}{N}\right), \tag{75}$$

$p_0$ being the initial pressure of the fluid into the porous material due to the atmospheric pressure and the gravity force. It is worth noticing that initial condition $(68)_1$ is not consistent with boundary condition (75), as a matter of fact $M \neq \overline{M}$ since $K_f \neq p_0$; moreover, when considering a water saturated porous material, which is generally the case in the consolidation problem, $K_f \gg p_0$ and therefore the term $\overline{M}b\epsilon$ in the boundary condition can be neglected with respect to the corresponding term $Mb\epsilon$ in the initial condition. This assumption implies Eq. (75) to reduce to Eq. $(68)_2$ in $x = 0$. Notice that the occurrence of $M \neq \overline{M}$ is crucial in modeling the consolidation phenomenon; if this were not the case no perturbation of the initial undrained conditions can arise in the Terzaghi model.

Boundary condition in $x = l$ should describe impermeability, which means that no fluid flux occurs through that surface; according to the classical statement of the Darcy law, such an assumption implies the pressure gradient to vanish: $\Delta p'(l, t) = 0$.

The weak point of the Terzaghi model is indeed in the description of the boundary layer close to the drained external surface; as it was already noticed a sharp discontinuity can be detected between the initial pressure, which is constant and non-zero, and the boundary condition in $x = 0$ which, on the other hand, prescribes drained conditions, i.e. a vanishing pressure.

### 7.2. The second gradient formulation of the consolidation problem

Consider now a one-dimensional formulation of the envisaged second gradient model, and in particular the one-dimensional governing equations coming from the overall balance law (13) and the linearized extended Darcy law (66). We embrace Terzaghi's hypothesis of neglecting inertia forces with respect to Darcy like drag actions: therefore both the overall bulk action $\mathbf{b}$ and that of the fluid $\mathbf{b}_f$ are assumed to vanish.

In particular we assume to deal with constitutive relations prescribed in terms of the overall potential $\Psi$ and reduce the linear constitutive relations stated in Eqs. (55)–(58) to the case of vanishing first gradient-second gradient couplings; in the following we therefore consider $\mathbf{\Pi}_{ss} = \mathbf{\Pi}_{sf} = 0$ and $\gamma_{fs} = \gamma_{ff} = 0$. This assumption corresponds to account for the effects of the wetting like coupling between the solid and the fluid inside a given RVE.

The perturbation of the pressure profile, with respect to its initial value, comes in this case from the solution of the following system of differential equations

$$(\lambda + 2\mu)\epsilon - bp_f^c - [\mathbb{K}_{ss} + \mathbb{M}k_{sf}(k_{sf} + b)]\epsilon'' - \frac{\mathbb{M}k_{sf}}{M}p_f^{c''} = -p^{ext}, \qquad (76)$$

$$\mathbb{M}\left[b\epsilon^{IV} + \frac{1}{M}(p_f^c)^{IV}\right] + \mathbb{M}k_{sf}\epsilon^{IV} - (p_f^c)^{II} - \alpha(\rho_f^0)^2\left[b\dot{\epsilon}^{II} + \frac{1}{M}(\dot{p}_f^c)^{II}\right] + a(\rho_f^0)^2\left(b\dot{\epsilon} + \frac{1}{M}\dot{p}_f^c\right) = 0; \qquad (77)$$

where $\alpha$ accounts also for the contribution due to the fourth order tensor $\mathbb{A}$ and $\mathbb{K}_{ss}$, $k_{sf}$ and $\mathbb{M}$ are the only non-vanishing scalar components of the corresponding second order tensors introduced in the constitutive relations – see Eqs. (55)–(58). Thus, in this case we can not deal just with a diffusion equation, but we need to determine the solution of a more complex problem.

Notice that neglecting inertia effects implies, as in the first gradient model, the balance equation of the porous material to reduce to a first integral.

The most interesting aspect of the second gradient formulation for the consolidation problem is, however, the statement of boundary and initial conditions: the boundary condition at $x = 0$ and the initial condition are in this case consistent. Because of the previous assumptions regarding the second gradient constitutive parameters, the initial condition still remains that of Eq. (68)$_1$. As for the classical consolidation theory, Eq. (9)$_2$ reads at $x = 0$ as a kind of drainage condition

$$\mathbf{t}_f = (n\sigma_f - \text{div}(n\mathbf{c}_f))\mathbf{n} \Rightarrow \phi p_f^c + \phi_0 \gamma_f' = \phi_0 p_0, \qquad (78)$$

where $\gamma_f$ is the only non-vanishing contribution to the hyper-stress vector and $p_0$ is the initial pressure of the fluid. In the frame of a linearized model Eq. (78) implies, according to the linearized constitutive relations (61) and (62),

$$\Delta p_f^c + \overline{M}b\epsilon - \overline{M}\frac{\phi_0}{p_0}\left[\mathbb{M}(k_{sf} + b)\epsilon'' + \frac{\mathbb{M}}{M}(\Delta p_f^c)''\right] = 0, \qquad (79)$$

where we use the same nomenclature as in Section 7.1. Analogous reasonings as those developed for the first gradient consolidation theory justify neglecting $\overline{M}b\epsilon$ in Eq. (79), with respect to $Mb\epsilon$, which appears in the initial condition; moreover, the second gradient terms of Eq. (79) are assumed to be of the same order of magnitude as the characteristic pressure $M$:

$$\frac{\mathbb{M}(k_{sf} + b)}{Ml^2} = O(1), \quad \frac{\mathbb{M}}{Ml^2} = O(1). \qquad (80)$$

The other boundary conditions require impermeability of the wall at $x = 0$, as in the first gradient model, and vanishing double forces (associated to the hyper-stress of the porous material and that one of the pores) at $x = 0$ and $x = l$.

Considering the extended formulation of the Darcy law (see Eq. (66)) and the constitutive laws (55)–(58) we get

$$-(\Delta p_f^c)' + \frac{\mathbb{M}}{M}(\Delta p_f^c)''' + \mathbb{M}(k_{sf} + b)\epsilon''' - \alpha(\rho_f^0)^2\left[b\dot{\epsilon}' + \frac{1}{M}(\Delta \dot{p}_f^c)'\right] = 0, \quad x = l, \qquad (81)$$

$$-\mathbb{M}\left[b\epsilon' + \frac{1}{M}(\Delta p_f^c)'\right] - \mathbb{M}k_{sf}\epsilon = 0, \quad x = 0, l \quad k_{sf}(\Delta p_f^c)' + [\mathbb{K}_{ss} + \mathbb{M}k_{sf}(k_{sf} + b)]\epsilon' = 0, \quad x = 0, l. \qquad (82)$$

Eqs. (76), (77) endowed with boundary conditions (79), (81) and (82) and initial condition (68)$_1$ define the differential problem.

### 7.3. A special solution of the second gradient theory of consolidation

In this section, we restrict our analysis to a particular second gradient model; we assume the following hypotheses for the constitutive parameters:

$$\mathbb{K}_{ss} + \mathbb{M}k_{sf}(k_{sf} + b) = 0, \quad \alpha = 0, \qquad (83)$$

which means that the constitutive relations for the hyper-stress $\gamma$ only depends on the gradient of the pore pressure but not on the strain gradient. The second assumption concerns the extended formulation of the Darcy law: in this case only the conservative additional effect associated to the second derivative of the pore hyper-stress is taken into account, whilst the Brinkman-like drag actions are neglected. Condition $(83)_1$ then allows for reducing Eqs. (76) and (77) to only one partial differential equation, which can be written in a suitable dimensionless form as

$$-\Gamma_4\Gamma_5 p^{\text{VI}} + (\Gamma_1 + b\Gamma_4)p^{\text{IV}} - p^{\text{II}} - \Gamma_2[(1+b^2\Gamma_3)\dot{p}^{\text{II}} - \Gamma_3\Gamma_5\dot{p}^{\text{IV}}] + [(1+b^2\Gamma_3)\dot{p} - \Gamma_3\Gamma_5\dot{p}^{\text{II}}] = 0, \tag{84}$$

where $p$ is the dimensionless pore pressure ($\Delta p_f^c = Mp$) and

$$\Gamma_1 := \frac{\mathbb{M}}{Ml^2}, \quad \Gamma_2 := \frac{\alpha}{al^2}, \quad \Gamma_3 := \frac{M}{\lambda + 2\mu}, \quad \Gamma_4 := \frac{\mathbb{M}(\mathsf{k}_{\text{sf}}+b)}{(\lambda+2\mu)l^2}, \quad \Gamma_5 := -\frac{\mathbb{M}\mathsf{k}_{\text{sf}}}{Ml^2}. \tag{85}$$

Spatial derivatives are made dimensionless with the characteristic length $l$ of the porous material, whilst time derivatives are made dimensionless with the characteristic time scale $\tau := al^2(\rho_f^0)^2/M$. Condition $(83)_2$ apparently implies $\Gamma_2 = 0$. Notice, Eq. (84) reduces to the classical Terzaghi equation (67) when $\Gamma_1 = \Gamma_2 = \Gamma_4 = \Gamma_5 = 0$.

Eq. (84) can be solved with the separation of variables technique, which means that we look for solutions of the form $p(x,t) = X(x)T(t)$, where $x$ and $t$ are the two dimensionless spatial and time variables. If we divide Eq. (84) by $[(1+b^2\Gamma_3)X - \Gamma_3\Gamma_5 X^{\text{II}}]T$ we get

$$\frac{-\Gamma_4\Gamma_5 X(x)^{\text{VI}} + (\Gamma_1 + b\Gamma_4)X(x)^{\text{IV}} - X(x)^{\text{II}}}{[(1+b^2\Gamma_3)X(x) - \Gamma_3\Gamma_5 X(x)^{\text{II}}]} = -\frac{\dot{T}(t)}{T(t)} = \text{const} =: -\lambda; \tag{86}$$

the two quantities, required to be equal, separately depend on $x$ and $t$, respectively. Eq. (86) easily implies the time function $T$ to be $T(t) = T_0\exp(\lambda t)$; the main goal is therefore to find solutions of the ordinary differential equation

$$-\Gamma_4\Gamma_5 X^{\text{VI}} + (\Gamma_1 + b\Gamma_4)X^{\text{IV}} - X^{\text{II}} + \lambda[(1+b^2\Gamma_3)X - \Gamma_3\Gamma_5 X^{\text{II}}] = 0, \tag{87}$$

endowed with the proper boundary conditions obtained from Eqs. (79), (81) and (82), which considering the separation of variables technique become:

$$-\Gamma_4\Gamma_5 X^{\text{IV}}(0) + (\Gamma_1 + b\Gamma_4)X^{\text{II}}(0) - X(0) = 0, \tag{88}$$
$$\Gamma_4\Gamma_5 X^{\text{III}}(0) - (\Gamma_1 + b\Gamma_4)X^{\text{I}}(0) = 0, \quad -\Gamma_3\Gamma_5 X^{\text{I}}(0) = 0, \tag{89}$$
$$-\Gamma_4\Gamma_5 X^{\text{V}}(1) + (\Gamma_1 + b\Gamma_4)X^{\text{III}}(1) - X^{\text{I}}(1) = 0, \tag{90}$$
$$\Gamma_4\Gamma_5 X^{\text{III}}(1) - (\Gamma_1 + b\Gamma_4)X^{\text{I}}(1) = 0, \quad -\Gamma_3\Gamma_5 X^{\text{I}}(1) = 0. \tag{91}$$

The spatial differential problem (87)–(91) is homogeneous with $\lambda$ as parameter; it is possible to prove that only when considering negative values of $\lambda$ non-trivial solutions can arise. In this case the differential problem looks like an eigenvalue problem. The eigenfunctions change when $\lambda$ belongs to different intervals of the real axis; in particular the characteristic equation

$$-\Gamma_4\Gamma_5\xi^3 + (\Gamma_1 + b\Gamma_4)\xi^2 - \xi + \lambda[(1+b^2\Gamma_3)X - \Gamma_3\Gamma_5\xi] = 0, \tag{92}$$

which is associated to Eq. (87) and is written in terms of the squared characteristic parameter $\xi$, can exhibit three real roots ($\xi_1(\lambda), \xi_2(\lambda) \in \mathbb{R}^+$ and $\xi_3(\lambda) \in \mathbb{R}^-$) or two complex roots ($\xi_1(\lambda) = \overline{\xi}_2(\lambda) \in \mathbb{C}$) and a negative real one ($\xi_3(\lambda) \in \mathbb{R}^-$). We do not present here the complete deduction of the eigenvalue analysis with respect to the constitutive parameters; conversely we directly state our claim on this topic: three open subsets $(0, \lambda_1)$, $(\lambda_1, \lambda_2)$ and $(\lambda_2, \infty)$ ($\lambda_1$ and $\lambda_2$ being positive constants) can be identified on the real axis in such a way that when $|\lambda|$ belongs to the first or the third interval we fall into case I (Eq. (92) admits two positive real roots and one negative real root), conversely when $|\lambda|$ belongs to the second interval we fall into case II:

case I $\xi_1(\lambda), \xi_2(\lambda) \in \mathbb{R}^+$ and $\xi_3(\lambda) \in \mathbb{R}^-$
$$X(x,\lambda) = K_1 \exp[\Psi_1(\lambda)x] + K_2 \exp[-\Psi_1(\lambda)x] + K_3 \exp[\Psi_2(\lambda)x] \\ + K_4 \exp[-\Psi_2(\lambda)x] + K_5 \cos[\Psi_3(\lambda)x] + K_6 \sin[\Psi_3(\lambda)x], \tag{93}$$

case II $\xi_1(\lambda) = \overline{\xi_2(\lambda)} \in \mathbb{C}$ and $\xi_3(\lambda) \in \mathbb{R}^-$
$$X(x,\lambda) = K_1 \exp[x \operatorname{Re}\Psi_1(\lambda)] \cos[x \operatorname{Im}\Psi_1(\lambda)] + K_2 \exp[x \operatorname{Re}\Psi_1(\lambda)] \sin[x \operatorname{Im}\Psi_1(\lambda)] \\ + K_3 \exp[-x \operatorname{Re}\Psi_1(\lambda)] \cos[x \operatorname{Im}\Psi_1(\lambda)] + K_4 \exp[-x \operatorname{Re}\Psi_1(\lambda)] \sin[x \operatorname{Im}\Psi_1(\lambda)] \\ + K_5 \cos[\Psi_2(\lambda)x] + K_6 \sin[\Psi_2(\lambda)x]. \tag{94}$$

With respect to the inner product in $H^3(\mathcal{I})(\mathcal{I} := (0,1))$

$$\langle f, g \rangle := (1+b^2\Gamma_3) \int_0^1 fg \, dx + [b\Gamma_3\Gamma_5 + (1+b^2\Gamma_3)(\Gamma_1 + b\Gamma_4)] \int_0^1 f^\mathrm{I} g^\mathrm{I} \, dx + \Gamma_5[\Gamma_4(1+b^2\Gamma_3) \\ + b\Gamma_3(\Gamma_1 + b\Gamma_4)] \int_0^1 f^\mathrm{II} g^\mathrm{II} \, dx + b\Gamma_3\Gamma_4\Gamma_5^2 \int_0^1 f^\mathrm{III} g^\mathrm{III} \, dx \tag{95}$$

the eigenfunctions associated to different eigenvalues are in both cases orthogonal to one another.

Here, we present some numerical results obtained for a consolidated clay saturated by water assuming the following values for the constitutive parameters:

| $E$ (GPa) | $\mu$ (GPa) | $N$ (GPa) | $\phi_0$ | $K_f$ (GPa) | $b$ |
|---|---|---|---|---|---|
| 4 | 1.43 | 133 | 0.4 | 2.15 | 0.9 |

(96)

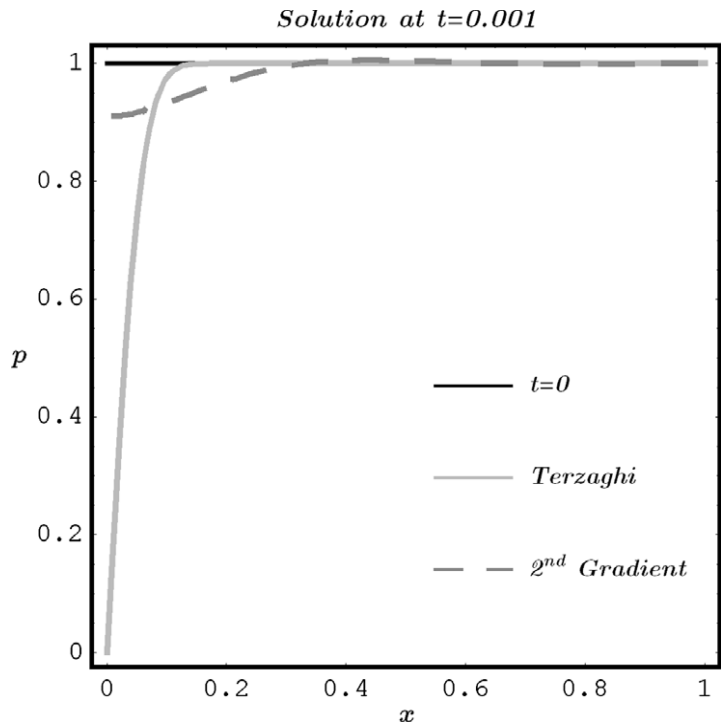

Fig. 1. Early time fluid pressure profiles obtained by classical consolidation theory (solid line) and second gradient theory (dashed line). The initial value of the fluid pressure is also depicted (black line). Dimensionless time value is explicitly indicated.

Pressure is made dimensionless so as to deal with a consolidating loading equal to unity:

$$p(x, 0^+) = \frac{bM}{K_u + \frac{4}{3}\mu} \frac{p^{ext}}{\bar{p}} = 1; \qquad (97)$$

moreover, second gradient constitutive parameters are introduced so as to capture a proper boundary layer effect close to the drained surface and the impermeable wall. In particular we compare the solution obtained with the second gradient theory of consolidation to the classical one due to Terzaghi at two characteristic time steps (see Figs. 1 and 2). The most interesting result of the comparison is the regularizing effect which characterizes the second gradient solution: the pressure profile does not fall to a vanishing pressure in $x = 0$ immediately after the initial state but progressively decreases to this limit value for $t \to \infty$. The obtained picture for the fluid pressure improves the information on the behaviour of the consolidating porous material close to the external surface; in particular we prove that the consolidating process does not proceed as quickly as Terzaghi's model predicts, but the fluid pressure initially hinders pore shrinkage enhancing, on the other hand, the effect of the consolidating pressure more inside the porous material. Reduction of the pore space to the equilibrium regime is therefore not instantaneous even in the vicinity of the external surface but wider than that predicted by Terzaghi's model. Indeed, a non-vanishing pore pressure is still admissible during consolidation at the external surface because of a non-vanishing gradient of pore hyper-stress; this is the main reason both for the reduction of the pressure jump close to the external surface and the enlargement of the boundary layer effect (see Fig. 1).

After a suitable time interval the second gradient model predicts a pore pressure increase at the impermeable wall ($x = l$) which was not described by Terzaghi's theory. This effect is mainly due the poor drainage of the region which is far from the consolidating surface; this is indeed a quite classical result when considering a porous slab of finite extent ($2a$) along $x$ and infinitely long in the other direction, sandwiched between two impermeable layers and drained at $x = \pm a$ (Mandel's problem, see Mandel, 1953), but it was never modeled with a one-dimensional theory of consolidation. As in the case of Mandel's theory this effect progressively decreases when time grows because of dissipative phenomena implied by the Darcy law; this effect is depicted in Fig. 2. These two specific features of the second gradient solution, with respect to the classical one, are really typical of second gradient models: actually it is the presence of boundary layers which mostly characterizes second gradient solutions, even in the absence of double forces. At early evolution times, i.e. when dissipation does not dominates the process, second gradient effects rise up close to the boundary of the domain; the characteristic size of this neighbourhood determines the intrinsic length of second gradient models. On the other hand when dissipation starts to dominate the time evolution process, i.e. when time goes to infinity, then the effect of second gradient becomes smaller and smaller.

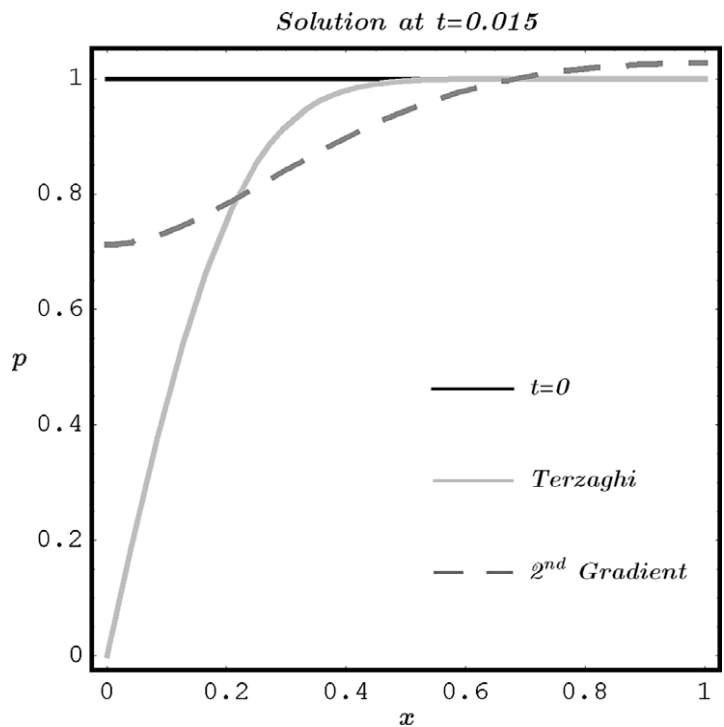

Fig. 2. Late time fluid pressure profile obtained by classical consolidation theory (solid line) and second gradient theory (dashed line). The initial value of the fluid pressure is also depicted (black line). A fluid overpressure at the impermeable wall arises. Dimensionless time value is explicitly indicated.

It is finally interesting to note that according to the representation formulas of the spatial function $X(x)$, see Eqs. (93) and (94), the second gradient solution does not exhibit the classical Gibbs effect associated to the attempt of recovering a Heavyside function (the initial condition for the fluid pressure) by means of Fourier series; therefore, the estimate of the effective solution at very early times obtained by truncation of the second gradient solution is much more refined than the Terzaghi one when accounting for the same number of eigenfunctions.

## 8. Concluding remarks

In this paper, the complete deduction of a second gradient poromechanical model is presented starting from a second gradient theory of mixtures; the main result consists of the identification of a properly defined macroscopic potential and consequently in the characterization of the second gradient extension of the classical Biot constitutive relations. Apparently, this is a completely phenomenological model and no explicit relation with the behaviour of the $a$th constituent over its characteristic domain has been sketched. In other words, only a macroscopic point of view has been adopted. However even in this frame some remarks can be stated. The key point is in recognizing the cases where the characteristic length of the heterogeneities can be compared at least with the thickness of the solid/fluid interface: for instance wetting and crack/pore opening phenomena. Further developments will be devoted to state a well grounded second gradient poromechanical model in the framework of micromechanics, on the basis of a proper definition of stresses and hyper-stresses.

In this paper, we prove that, for a second gradient behaviour of the solid and the fluid constituent, i.e. prescribing an external work rate which depends on tractions and double forces, see Eq. (16), and if the extended Cauchy stress theorem holds true, see Eq. (7), then suitable macroscopic Helmholtz free energies ($\Psi$ and $\Psi_s$) can be determined not only as functions of strain and fluid mass (porosity) but also of strain and fluid mass (porosity) gradient. This result is obtained by considering the thermodynamic restrictions coming from the first and the second principle of thermodynamics, and in particular from the extended formulation of the classical Causius–Duhem inequality. According to this result, we prove that a natural extension of the Biot model can be formulated in order to generalize the classical theory of porous media to the case of second gradient materials; the introduced additional extensive variables, i.e. gradient of strain and gradient of fluid mass (porosity), play an analogous role as strain and fluid mass (porosity) in the classical model. The corresponding intensive parameters, i.e. the overall hyper-stress and the pure fluid hyper-stress, are constitutively prescribed in terms of first and second gradient deformation variables; moreover, the classical constitutive relations for the overall stress and the fluid mass content (porosity) are corrected by additional second gradient terms (see e.g. Eqs. (55)–(58) and Eqs. (59)–(62) which are valid in the linearized case). According to the aforementioned thermodynamic restrictions we prove that the classical Darcy law becomes a Brinkman-like equation involving the fluid pressure and the velocity of the fluid relative to the solid.

In the last section a numerical example is developed where the corrections due to second gradients are investigated for the classical one-dimensional consolidation problem stated by Terzaghi for a homogeneous porous material. Considering the fluid pressure profiles, the differences concern the boundary layers which now can be detected in the neighbourhood of the consolidating surface (the external surface – $x = 0$) and the impermeable wall (the internal surface – $x = l$). In particular, the second gradient removes the singular behaviour due to the initial – boundary condition discontinuity at $x = 0$ and implies a non-trivial dilatant behaviour close to the impermeable wall before dissipative diffusion starts to govern the process.

## Appendix A. Dissipation

In this Appendix, we illustrate how Eqs. (37)–(39) can be recovered starting from the general formulation of the second principle (35) by means of Eq. (29) which gives the representation formula for the overall Helmholtz free energy, and Eq. (32) which provides the local form of the first principle of thermodynamics pulled back into the reference configuration of the skeleton.

Start from Eq. (35) and replace the time derivative of the overall Lagrangean internal energy $E$ with its representation formula coming from the first principle – see Eq. (32)

$$[S - C^{-1}((\nabla_0 C)\gamma)C^{-1}] \cdot \frac{d\Delta}{dt} + C^{-1} \otimes \gamma \cdot \frac{d\nabla_0\Delta}{dt} - \text{div}_0(e_f M + Q) - \text{div}_0\left[\frac{1}{m_f}\text{div}_0(n\gamma_f)M + \frac{p_f}{\rho_f}M\right]$$
$$+ \text{div}_0\left\{\frac{1}{\text{tr}\mathbf{I}}\left[\frac{\gamma_f}{J}\cdot\left(\nabla_0\frac{1}{\rho_f} - \frac{1}{n\rho_f}\nabla_0 n\right)M\right] + \text{dev}\left[\frac{n\gamma_f}{J}\otimes\nabla_0\left(\frac{1}{n\rho_f}\right)\right]M + \frac{\gamma_f}{J\rho_f}\text{div}_0 M\right\}$$
$$+ \text{div}_0\left(\frac{1}{J\rho_f}\text{dev}S_f^\top M\right) + T\text{div}_0\left(s_f M + \frac{Q}{T}\right) - S\frac{dT}{dt} - \frac{d\Psi}{dt} \geq 0, \tag{98}$$

consider the time derivative of the overall energy $\Psi$, as defined by Eq. (29), and recall in particular the introduced variation $d_n$

$$\frac{d\Psi}{dt} = \frac{d\Psi_s}{dt} + \psi_f\frac{dm_f}{dt} + m_f\left(\frac{d_n e_f}{dt} - s_f\frac{dT}{dt} - T\frac{ds_f}{dt}\right) \tag{99}$$

substitute it into Eq. (98) and develop some calculations in the formula so as to obtain

$$[S - C^{-1}((\nabla_0 C)\gamma)C^{-1}]\cdot\frac{d\Delta}{dt} + C^{-1}\otimes\gamma\cdot\frac{d\nabla_0\Delta}{dt} + \text{div}_0\left(\frac{1}{J\rho_f}\text{dev}S_f^\top M\right) - \nabla_0\left(e_f + \frac{p_f}{\rho_f}\right)\cdot M$$
$$- \nabla_0\left(\frac{1}{m_f}\text{div}_0(n\gamma_f)\right)\cdot M - \left(e_f + \frac{p_f}{\rho_f} + \frac{1}{m_f}\text{div}_0(n\gamma_f)\right)\text{div}_0 M + \text{dev}\left[\nabla_0\left(\frac{1}{n\rho_f}\right)\otimes\frac{n\gamma_f}{J}\right]\cdot\nabla_0 M$$
$$+ \text{div}_0\left\{\text{dev}\left[\nabla_0\left(\frac{1}{n\rho_f}\right)\otimes\frac{n\gamma_f}{J}\right]\right\}\cdot M + \frac{1}{\text{tr}\mathbf{I}}\frac{\gamma_f}{J}\cdot\left(\nabla_0\frac{1}{\rho_f} - \frac{1}{n\rho_f}\nabla_0 n\right)\text{div}_0 M + \frac{1}{\text{tr}\mathbf{I}}\nabla_0\left[\frac{\gamma_f}{J}\cdot\left(\nabla_0\frac{1}{\rho_f} - \frac{1}{n\rho_f}\nabla_0 n\right)\right]\cdot M$$
$$+ \text{div}_0\left(\frac{\gamma_f}{J\rho_f}\right)\text{div}_0 M + \frac{\gamma_f}{J\rho_f}\cdot\nabla_0\text{div}_0 M + Ts_f\text{div}_0 M + \nabla_0(Ts_f)\cdot M - s_f M\cdot\nabla_0 T - \frac{1}{T}\nabla_0 T\cdot Q - S_s\frac{dT}{dt} - \frac{d\Psi_s}{dt}$$
$$- \psi_f\frac{dm_f}{dt} - m_f\frac{d_n e_f}{dt} + m_f T\frac{ds_f}{dt} \geq 0. \tag{100}$$

Split the fluid pressure $p_f$ into a conservative and a dissipative contribution, say $p_f^c$ and $p_f^d$, respectively, compute the time derivative of $e_f$ according to Eq. (25) and account for the following identity

$$\frac{\partial e_f}{\partial \nabla(n\rho_f)}\cdot\frac{d_n}{dt}\nabla(n\rho_f) = -L\cdot[F^{-\top}\nabla_0(n\rho_f)]\otimes\frac{\partial e_f}{\partial \nabla(n\rho_f)} + F^{-1}\frac{\partial e_f}{\partial \nabla(n\rho_f)}\cdot\frac{d_n}{dt}\nabla_0(n\rho_f)$$
$$= -\frac{d\Delta}{dt}\cdot[C^{-1}\nabla_0(n\rho_f)]\otimes\left[F^{-1}\frac{\partial e_f}{\partial \nabla(n\rho_f)}\right] + F^{-1}\frac{\partial e_f}{\partial \nabla(n\rho_f)}\cdot\frac{d_n}{dt}\nabla_0(n\rho_f), \tag{101}$$

$L := \dot F F^{-1}$ being the gradient of the velocity of the solid. Considering the statement of the balance of mass of the fluid pulled back in the reference configuration of the skeleton $(dm_f/dt + \text{div}_0 M = 0)$ and bearing in mind the identity

$$\frac{d}{dt}\left[\nabla_0\left(\frac{m_f}{\rho_f}\right)\right] = \frac{d}{dt}\left(\frac{1}{\rho_f}\nabla_0 m_f + m_f\nabla_0\frac{1}{\rho_f}\right) \tag{102}$$
$$= \frac{d}{dt}\left(\frac{1}{\rho_f}\right)\nabla_0 m_f + \frac{1}{\rho_f}\frac{d}{dt}(\nabla_0 m_f) + m_f\frac{d}{dt}\left(\nabla_0\frac{1}{\rho_f}\right) + \left(\nabla_0\frac{1}{\rho_f}\right)\frac{dm_f}{dt}$$

Eq. (100) reads, according to Eqs. (26)–(28), like this

$$\left\{ S - \mathbf{C}^{-1}((\nabla_0 \mathbf{C})\boldsymbol{\gamma})\mathbf{C}^{-1} - [\mathbf{C}^{-1}\nabla_0(n\rho_f)] \otimes \frac{\boldsymbol{\gamma}_f}{\rho_f} \right\} \cdot \frac{d\boldsymbol{\Delta}}{dt} + \mathbf{C}^{-1} \otimes \boldsymbol{\gamma} \cdot \frac{d\nabla_0 \boldsymbol{\Delta}}{dt} + \mathrm{div}_0 \left( \frac{1}{J\rho_f} \mathrm{dev} S_f^\top \mathbf{M} \right)$$

$$- \nabla_0 \left( e_f + \frac{p_f}{\rho_f} \right) \cdot \mathbf{M} - \nabla_0 \left( \frac{1}{m_f} \mathrm{div}_0(n\boldsymbol{\gamma}_f) \right) \cdot \mathbf{M} + \mathrm{div}_0 \left\{ \mathrm{dev} \left[ \nabla_0 \left( \frac{1}{n\rho_f} \right) \otimes \frac{n\boldsymbol{\gamma}_f}{J} \right] \right\} \cdot \mathbf{M}$$

$$+ \mathrm{dev} \left[ \nabla_0 \left( \frac{1}{n\rho_f} \right) \otimes \frac{n\boldsymbol{\gamma}_f}{J} \right] \cdot \nabla_0 \mathbf{M} - \frac{1}{\mathrm{tr}\mathbf{I}} \nabla_0 \left[ \frac{\boldsymbol{\gamma}_f}{J} \cdot \left( \frac{1}{\rho_f^2} \nabla_0 \rho_f + \frac{\nabla_0 n}{n\rho_f} \right) \right] \cdot \mathbf{M} - \frac{p_f^d}{\rho_f} \mathrm{div}_0 \mathbf{M} + T\nabla_0 s_f \cdot \mathbf{M}$$

$$+ \frac{1}{\rho_f} \left( p_f^c + \frac{1}{\phi} \nabla_0 n \cdot \boldsymbol{\gamma}_f - \boldsymbol{\gamma}_f \cdot \nabla_0 \frac{1}{J} \right) \frac{dm_f}{dt} + \frac{\boldsymbol{\gamma}_f}{J} \cdot \left[ \frac{d}{dt}\left(\frac{1}{\rho_f}\right) \nabla_0 m_f + m_f \frac{d}{dt} \nabla_0 \left(\frac{1}{\rho_f}\right) \right] - \frac{\boldsymbol{\gamma}_f}{J} \cdot \frac{d}{dt} \nabla_0 \phi$$

$$+ \frac{1}{\mathrm{tr}\mathbf{I}} \frac{\boldsymbol{\gamma}_f}{J} \cdot \left( \frac{1}{\rho_f^2} \nabla_0 \rho_f + \frac{\nabla_0 n}{n\rho_f} \right) \frac{dm_f}{dt} + m_f \left[ \left( p_f^c + \left(1 + \frac{1}{\mathrm{tr}\mathbf{I}}\right) \frac{\boldsymbol{\gamma}_f}{J} \cdot \frac{\nabla_0(n\rho_f)}{n\rho_f} \right) \frac{d}{dt}\left(\frac{1}{\rho_f}\right) + \frac{n\boldsymbol{\gamma}_f}{J(n\rho_f)^2} \cdot \frac{d_n}{dt} \nabla_0(n\rho_f) \right]$$

$$- \frac{1}{T} \nabla_0 T \cdot \mathbf{Q} - S_s \frac{dT}{dt} - \frac{d\Psi_s}{dt} \geqslant 0. \tag{103}$$

Considering the equalities

$$\frac{1}{\mathrm{tr}\mathbf{I}} \frac{\boldsymbol{\gamma}_f}{J} \cdot \left( \frac{1}{\rho_f^2} \nabla_0 \rho_f + \frac{\nabla_0 n}{n\rho_f} \right) \frac{dm_f}{dt} = \frac{1}{\mathrm{tr}\mathbf{I}} \frac{\boldsymbol{\gamma}_f}{J} \cdot \frac{1}{\rho_f} \left( \frac{1}{n\rho_f} \nabla_0(n\rho_f) \right) \frac{dm_f}{dt}$$

$$\left( p_f^c + \frac{1}{\mathrm{tr}\mathbf{I}} \frac{\boldsymbol{\gamma}_f}{J} \cdot \frac{1}{n\rho_f} \nabla_0(n\rho_f) \right) \left( \frac{1}{\rho_f} \frac{dm_f}{dt} + m_f \frac{d}{dt}\left(\frac{1}{\rho_f}\right) \right) = \left( p_f^c + \frac{1}{\mathrm{tr}\mathbf{I}} \frac{\boldsymbol{\gamma}_f}{J} \cdot \frac{1}{n\rho_f} \nabla_0(n\rho_f) \right) \frac{d\phi}{dt}$$

we finally get the complete dissipation formula

$$\left\{ S - \mathbf{C}^{-1}((\nabla_0 \mathbf{C})\boldsymbol{\gamma})\mathbf{C}^{-1} - [\mathbf{C}^{-1}\nabla_0(n\rho_f)] \otimes \frac{\boldsymbol{\gamma}_f}{\rho_f} \right\} \cdot \frac{d\boldsymbol{\Delta}}{dt} + \mathbf{C}^{-1} \otimes \boldsymbol{\gamma} \cdot \frac{d\nabla_0 \boldsymbol{\Delta}}{dt} + \mathrm{div}_0 \left( \frac{1}{J\rho_f} \mathrm{dev} S_f^\top \mathbf{M} \right)$$

$$- \nabla_0 \left( e_f + \frac{p_f}{\rho_f} \right) \cdot \mathbf{M} - \nabla_0 \left( \frac{1}{m_f} \mathrm{div}_0(n\boldsymbol{\gamma}_f) \right) \cdot \mathbf{M} + \mathrm{div}_0 \left\{ \mathrm{dev} \left[ \nabla_0 \left( \frac{1}{n\rho_f} \right) \otimes \frac{n\boldsymbol{\gamma}_f}{J} \right] \right\} \cdot \mathbf{M}$$

$$+ \mathrm{dev} \left[ \nabla_0 \left( \frac{1}{n\rho_f} \right) \otimes \frac{n\boldsymbol{\gamma}_f}{J} \right] \cdot \nabla_0 \mathbf{M} - \frac{1}{\mathrm{tr}\mathbf{I}} \nabla_0 \left[ \frac{\boldsymbol{\gamma}_f}{J} \cdot \left( \frac{1}{\rho_f^2} \nabla_0 \rho_f + \frac{\nabla_0 n}{n\rho_f} \right) \right] \cdot \mathbf{M} - \frac{p_f^d}{\rho_f} \mathrm{div}_0 \mathbf{M} + T\nabla_0 s_f \cdot \mathbf{M}$$

$$+ \left( p_f^c + \frac{1}{\mathrm{tr}\mathbf{I}} \frac{\boldsymbol{\gamma}_f}{J} \cdot \frac{1}{n\rho_f} \nabla_0(n\rho_f) - \frac{\boldsymbol{\gamma}_f}{J} \cdot \phi \nabla_0 \frac{1}{\phi} \right) \frac{d\phi}{dt} - \frac{\boldsymbol{\gamma}_f}{J} \cdot \frac{d}{dt} \nabla_0 \phi - \frac{1}{T} \nabla_0 T \cdot \mathbf{Q} - S_s \frac{dT}{dt} - \frac{d\Psi_s}{dt} \geqslant 0.$$

## References

Allaire, G., 1997. One phase Newtonian flow. In: Hornung, U. (Ed.), Homogenization and Porous Media. Springer-Verlag, New York, pp. 45–68.

Anderson, D.M., McFadden, G.B., Wheeler, A.A., 1998. Diffuse-interface methods in fluid mechanics. Annu. Rev. Fluid Mech. 30, 139–165.

Biot, M.A., 1941. General theory of three-dimensional consolidation. J. Appl. Phys. 12, 155–164.

Brinkman, H.C., 1947. A calculation of the viscous force exerted by a flowing fluid on a dense swarm of particles. Appl. Sci. Res. A 1, 27–34.

Cahn, J.W., Hilliard, J.E., 1959. Free energy of a non-uniform system. J. Chem. Phys. 31, 688–699.

Casal, P., 1972. La théorie du second gradient et la capillarité. C.R. Acad. Sci. Paris Série A 274, 1571–1574.

Coleman, B.D., Noll, W., 1963. The thermodynamics of elastic material with heat conduction and viscosity. Arch. Rational Mech. Anal. 13, 167–178.

Collin, F., Chambon, R., Charlier, R., 2006. A finite element method for poro mechanical modelling of geotechnical problems using local second gradient models. Int. J. Num. Meth. Engng. 65, 1749–1772.

Coussy, O., Dormieux, L., Detournay, E., 1998. From mixture theory to Biot's approach for porous media. Int. J. Solids Struct. 35 (34–35), 4619–4635.

Coussy, O., 2004. Poromechanics. John Wiley and Sons, Chichester.

de Boer, R., 1996. Highlights in the historical development of the porous media theory: toward a consistent macroscopic theory. Appl. Mech. Rev. 49 (4), 201–261.

de Gennes, P.G., 1985. Wetting: statics and dynamics. Rev. Modern Phys. 57 (3), 827–863.

Degiovanni, M., Marzocchi, A., Musesti, A., 2006. Edge-force densities and second-order powers. Ann. Matematica 185, 81–103.

dell'Isola, F., Gouin, H., Rotoli, G., 1996. Nucleation of spherical shell-like interfaces by second gradient theory: numerical simulations. Eur. J. Mech. B/Fluids 15, 545–568.

dell'Isola, F., Seppecher, P., 1997. Edge contact forces and quasi balanced power. Meccanica 32, 33–52.

dell'Isola, F., Guarascio, M., Hutter, K., 2000. A variational approach for the deformation of a saturated porous solid. A second gradient theory extending Terzaghi's effective stress principle. Arch. Appl. Mech. 70, 323–337.

dell'Isola, F., Sciarra, G., Batra, R.C., 2003. Static Deformations of a Linear Elastic Porous Body filled with an Inviscid Fluid. J. Elasticity 72, 99–120.

Dormieux, L., Coussy, O., de Buhan, P., 1991. Modélisation d'un milieu polyphasique par la méthode des puissances virtuelles. C. R. Acad. Sci. Paris Série II 313, 863–868.

Dormieux, L., Ulm, F.J. (Eds.), 2005. Applied micromechanics of porous materials series CISM n. 480. Springer-Verlag, Wien, New York.

Drugan, W.J., Willis, J.R., 1996. A micromechanics-based nonlocal constitutive equation and estimates of representative volume element size for elastic component. J. Mech. Phys. Solids 44 (4), 497–524.

Gavrilyuk, S., Saurel, R., 2002. Mathematical and numerical modeling of two-phase compressible flows with micro-inertia. J. Comput. Phys. 175, 326–360.

Germain, P., 1973. La méthode des puissances virtuelles en mécanique des milieux continus. Journal de Mécanique 12 (2), 235–274.

Gurtin, M.E., 1981. An Introduction to Continuum Mechanics. Academic Press, Boston.

Gurtin, M.E., Polignone, D., Vinals, J., 1996. Two-phase binary fluids and immiscible fluids described by an order parameter. Math Models Meth. Appl. Sci. 6, 815–831.

Lee, H.G., Lowengrub, J.S., Goodman, J., 2002. Modeling pinchoff and reconnection in a Hele–Shaw cell. I. The models and their calibration. Phys. Fluids 14, 492–513.

Lowengrub, J., Truskinovsky, L., 1998. Quasi-incompressible Cahn–Hilliard fluids and topological transitions. Proc. R. Soc. Lond. A 454, 2617–2654.

Mandel, J., 1953. Consolidation des sols (étude mathématique). Géotechnique 3, 287–299.

Mindlin, R.D., 1964. Micro-structure in linear elasticity. Arch. Rat. Mech. Anal. 16, 51–78.

Modica, L., 1987a. Theory of phase transitions and minimal interface criterion. Arch. Rat. Mech. Anal. 98, 123–142.

Modica, L., 1987b. Gradient theory of phase transitions with boundary contact energy. Ann. Inst. Henri Poincaré Analyse non linéaire 4 (5), 487–512.

Sciarra, G., dell'Isola, F., Hutter, K., 2001. A solid–fluid mixture model allowing for solid dilatation under external pressure. Continuum Mech. Thermodyn. 13, 287–306.

Sciarra, G., 2001. Modelling of a fluid flux through a solid deformable matrix. PhD Thesis University of Rome "La Sapienza" and Université de Toulon et du Var.

Seppecher, P., 1987. Etude d'une modelisation des zones capillaires fluides: interfaces et lignes de contact. Thèse Université. Paris VI.

Seppecher, P., 1993. Equilibrium of a Cahn–Hilliard fluid on a wall: influence of the wetting properties of the fluid upon the stability of a thin liquid film. Eur. J. Mech. B/Fluids 12, 169–184.

Seppecher, P., 1996. Moving contact lines in the Cahn–Hilliard theory. Int. J. Eng. Sci. 34 (9), 977–992.

Suiker, A.S.J., de Borst, R., Chang, C.S., 2001. Micro-mechanical modelling of granular material. Part I: Derivation of a second-gradient micro-polar constitutive theory. Acta Mech. 149, 161–180.

von Terzaghi, K., 1946. Theoretical Soil Mechanics. John Wiley and Sons.

Toupin, R.A., 1962. Elastic materials with couple-stresses. Arch. Rat. Mech. Anal. 11, 385–414.

Toupin, R.A., 1964. Theories of elasticity with couple-stress. Arch. Rat. Mech. Anal. 17, 85–112.

Truesdell, C.A., 1991. A First Course in Rational Continuum Mechanics, second ed. Academic Press, Boston.